\documentclass[12pt]{article}

\usepackage{amssymb}
\usepackage{amsmath}
\usepackage{amsfonts}

\newcommand{\R}{\mathbb{R}}
\newcommand{\C}{\mathbb{C}}

\newcommand{\be}{\begin{equation}}
\newcommand{\bea}{\begin{eqnarray}}
\newcommand{\eea}{\end{eqnarray}}
\newcommand{\nn}{\nonumber}
\newcommand{\kt}{\rangle}
\newcommand{\br}{\langle}

\newcommand{\ed}{\end{document}}
\newcommand{\bbr}{\br\!\br}
\newcommand{\kkt}{\kt\!\kt}
\newcommand{\pbr}{\prec}
\newcommand{\pkt}{\succ}
\newcommand{\ppbr}{\prec\!\!\!\prec}
\newcommand{\ppkt}{\succ\!\!\!\succ}
\newcommand{\cbr}{(\!(}
\newcommand{\ckt}{)\!)}
\begin{document}

\title{Quantum Mechanics of Klein-Gordon-Type Fields
and Quantum Cosmology}
\author{\\
Ali Mostafazadeh
%\thanks{E-mail address: amostafazadeh@ku.edu.tr}\\
\\ Department of Mathematics, Ko\c{c} University,\\
Rumelifeneri Yolu, 34450 Sariyer,\\
Istanbul, Turkey\\ \\
amostafazadeh@ku.edu.tr}
\date{ }
\maketitle
\newpage

\baselineskip=24pt
\textheight = 22.5cm \topskip = -1.5cm \topmargin = -1.5cm

\begin{abstract}
With a view to address some of the basic problems of quantum
cosmology, we formulate the quantum mechanics of the solutions of
a Klein-Gordon-type field equation: $(\partial_t^2+D)\psi(t)=0$,
where $t\in\R$ and $D$ is a positive-definite operator acting in a
Hilbert space $\tilde{\cal H}$. In particular, we determine all
the positive-definite inner products on the space ${\cal H}$ of
the solutions of such an equation and establish their physical
equivalence. This specifies the Hilbert space structure of ${\cal
H}$ uniquely. We use a simple realization of the latter to
construct the observables of the theory explicitly. The field
equation does not fix the choice of a Hamiltonian operator unless
it is supplemented by an underlying classical system and a
quantization scheme supported by a correspondence principle. In
general, there are infinitely many choices for the Hamiltonian
each leading to a different notion of time-evolution in ${\cal
H}$. Among these is a particular choice that generates
$t$-translations in ${\cal H}$ and identifies $t$ with time
whenever $D$ is $t$-independent. For a $t$-dependent $D$, we show
that regardless of the choice of the inner product the
$t$-translations do not correspond to unitary evolutions in ${\cal
H}$, and $t$ cannot be identified with time. We apply these ideas
to develop a formulation of quantum cosmology based on the
Wheeler-DeWitt equation for a Friedman-Robertson-Walker model
coupled to a real scalar field with an arbitrary positive
confining potential. In particular, we offer a complete solution
of the Hilbert space problem, construct the observables, use a
position-like observable to introduce the wave functions of the
universe (which differ from the Wheeler-DeWitt fields),
reformulate the corresponding quantum theory in terms of the
latter, reduce the problem of the identification of time to the
determination of a Hamiltonian operator acting in $L^2(\R)\oplus
L^2(\R)$, show that the factor-ordering problem is irrelevant for
the kinematics of the quantum theory, and propose a formulation of
the dynamics. Our method is based on the central postulates of
nonrelativistic quantum mechanics, especially the quest for a
genuine probabilistic interpretation and a unitary Schr\"odinger
time-evolution. It generalizes to arbitrary minisuperspace
(spatially homogeneous) models and provides a way of unifying the
two main approaches to the canonical quantum cosmology based on
these models, namely quantization before and after imposing the
Hamiltonian constraint.
\end{abstract}
%\vspace{2mm}

%\textheight = 22.5cm \topskip = -1.5cm \topmargin = -1.5cm
\baselineskip=18pt

\newpage
\section{Introduction}\label{s1}

The problem of applying the machinery of nonrelativistic quantum
mechanics to Klein-Gordon fields has been a subject of interest
since late 1920s. It was this problem that led Dirac to the
discovery of the wave equation for the electron and the
formulation of the method of second quantization. These form the
main ingredients of the modern theories of elementary particle
physics. The very same problem also arises in canonical quantum
gravity \cite{bryce-1} and in particular quantum cosmology where
the trick of considering a first order field equation such as
Dirac's or using the method of second quantization does not
provide a satisfactory description \cite{unruh-wald,isham,carlip}.
It is ironic that despite its enormous impact on the formulation
and resolution of a number of fundamental problems of theoretical
physics, a satisfactory solution of this problem has been out of
reach. The purpose of this article is two-fold. First, it aims at
providing a complete solution of the above-mentioned problem for
the class of linear field equations of the form
    \be
    \ddot\psi(t)+D\psi(t)=0,
    \label{kgt}
    \end{equation}
where $t\in\R$, a dot denotes a $t$-derivative, and $D:\tilde{\cal
H}\to\tilde{\cal H}$ is a possibly $t$-dependent Hermitian
operator acting in a Hilbert space $\tilde{\cal H}$ from which one
selects the values $\psi(t)$ of the field $\psi$, alternatively
the initial values of the field $\psi_0$ and its $t$-derivative
$\dot\psi_0$ at an initial value $t_0$ of $t$. Second, it employs
the resulting theory to devise a formulation of the minisuperspace
quantum cosmology that allows for a genuine probabilistic
interpretation and a unitary Schr\"odinger time-evolution.

Equation~(\ref{kgt}) is a simple generalization of the free
Klein-Gordon equation, for the latter corresponds to the choice:
    \be
    t={\rm time},~~~~~~~~~~~~~
    \tilde{\cal H}=L^2(\R^3),~~~~~~~~~~~~~~
    D=-\nabla^2+\mu^2,
    \label{kg}
    \end{equation}
where $\mu:=mc/\hbar$ and $m$ is the mass of the Klein-Gordon
field. Therefore, following Ref.~\cite{cqg}, we call (\ref{kgt})
(respectively its solutions $\psi$) a {\em Klein-Gordon-type field
equation} (respectively {\em Klein-Gordon-type fields}). The
Wheeler-DeWitt equation for a number of minisuperspace
cosmological models \cite{misner} also provides a family of
Klein-Gordon-type equations. A well-known example is the
Wheeler-DeWitt equation associated with a
Friedman-Robertson-Walker (FRW) model coupled to a real scalar
field \cite{kaup-vitello,blyth-isham,isham,page,wiltshire}:
    \be
    \left[ -\frac{\partial^2}{\partial\alpha^2}+
    \frac{\partial^2}{\partial\varphi^2}+
    \kappa\,e^{4\alpha}-e^{6\alpha}V(\varphi)\right]\,
    \psi(\alpha,\varphi)=0,
    \label{wdw-0}
    \end{equation}
where $\alpha:=\ln a$, $a$ is the scale factor, $V=V(\varphi)$ is
a real-valued potential for the field $\varphi$, $\kappa=-1,0,1$
determines whether the FRW model describes an open, flat, or
closed universe, respectively, and we have chosen a particularly
simple factor-ordering and the natural units
\cite{hawking-page,wiltshire}. We can write the Wheeler-DeWitt
equation~(\ref{wdw-0}) in the form~(\ref{kgt}), if we identify
$\alpha$ with the variable $t$, set $\tilde{\cal H}=L^2(\R)$ and
$D=-\partial^2_\varphi+e^{6\alpha}\,V(\varphi)-\kappa\,e^{4\alpha}$.
Clearly $D$ is a Hermitian operator acting in
$L^2(\R)$.\footnote{We will avoid identifying $\alpha=t$ with a
physical time variable. We will show that this choice violates
unitarity!}

The identification of the Wheeler-DeWitt equation for various
minisuperspace models with certain Klein-Gordon-type field
equations marks the significance of devising a genuine quantum
mechanical treatment of a general Klein-Gordon-type field. This
requires the identification of an appropriate Hilbert space (a
vector space endowed with a complete positive-definite inner
product) that determines the kinematics and a Hermitian
Hamiltonian operator that governs the dynamics of the
corresponding theory.

The natural choice for the vector space structure of the Hilbert
space is the solution space of the corresponding Klein-Gordon-type
field equation \cite{crnkovic-witten}. Endowing this vector space
with an `appropriate' positive-definite inner product is a more
difficult task. In Ref.~\cite{cqg}, we constructed a class of
positive-definite inner products on the solution space ${\cal H}$
of a Klein-Gordon-type equation. In this article, we give a direct
argument clarifying the extent of the generality of the results of
\cite{cqg}. In particular, we address the problem of finding the
most general positive-definite inner product on ${\cal H}$ that
turns it into a Hilbert space. We show that the corresponding
Hilbert spaces are physically equivalent. Furthermore, for various
choices of the inner product (various realizations of the Hilbert
space structure), we discuss in great detail the nature and
construction of the possible Hamiltonian operators and other
observables of the corresponding quantum systems. The choice of a
Hamiltonian operator determines what we mean by `time-evolution.'
We show that, similarly to nonrelativistic quantum mechanics, this
choice is by no means unique unless one identifies a corresponding
classical system and employs a quantization scheme (with specific
operator-ordering prescription in canonical quantization or choice
of the `measure' in the path-integral quantization).

A direct application of our general results in minisuperspace
quantum cosmology provides a way of decoupling the Hilbert space
problem and the problem of time. It leads to a resolution of the
former and suggests how one should approach the latter. It further
shows that the factor-ordering ambiguity associated with the
Wheeler-DeWitt equation does not affect the kinematical structure
of the corresponding quantum theory. Another remarkable
consequence of our method is that in a sense it unifies the two
main approaches to the canonical quantization of gravity, namely
quantization before and after imposing the constraints
\cite{isham}.

The article is organized as follows. In Section~2, we provide a
brief review of the results of Ref.~\cite{cqg}. In Section~3, we
construct the most general positive-definite inner product on the
solution space of a Klein-Gordon-type equation. In Section~4, we
study a special class of Klein-Gordon-type fields, qualified as
being stationary, give various equivalent descriptions of the
quantum mechanics for this class, explore the corresponding
Hamiltonians, and construct the observables. In Section~5, we
extend the results of Section~4 to general (nonstationary)
Klein-Gordon-type fields. In Section~6, we demonstrate the
application of our general results in the study of the quantum
mechanics on the solution space of the equation of motion for a
classical (possibly time-dependent) harmonic oscillator. A
particular example of the latter is the Wheeler-DeWitt equation
for a FRW cosmological model with a cosmological constant. In
section~7, we consider the FRW models coupled to a real scalar
field with an arbitrary positive confining potential $V$. We
develop the corresponding quantum cosmology, i.e., construct a
positive-definite inner product on the physical Hilbert space of
the solutions of the Wheeler-DeWitt equation~(\ref{wdw-0}), define
a set of basic observables, introduce a position-like basis for
the Hilbert space and use it to define a wave function $f$
associated with every Wheeler-DeWitt field $\psi$. Using the fact
that in our approach Wheeler-DeWitt fields $\psi$ are treated as
vectors belonging to the abstract Hilbert space ${\cal H}$ of the
theory, we explain the conceptual difference between the functions
$\psi(\alpha,\varphi)$ appearing in the Wheeler-DeWitt
equation~(\ref{wdw-0}) and the wave functions $f$ that assign the
coefficients of the Wheeler-DeWitt fields $\psi$ in the
position-like basis. This suggests that it is the wave functions
$f$ that should be identified with the `wave functions of the
universe' not the functions $\psi(\alpha,\varphi)$. We give a
formulation of the kinematics and the dynamics in terms of these
wave functions, and reduce the problem of the identification of
time to the issue of selecting a Hamiltonian operator (a linear
Hermitian operator) acting in the space $L^2(\R)\oplus L^2(\R)$ of
the wave functions $f$. Following Dirac's canonical quantization
program, we then argue that the Hamiltonian operator is to be
obtained by quantizing a classical Hamiltonian that governs the
classical `dynamics' of the system after imposing the classical
constraint. This requires the identification of the position-like
operator with the quantum analog of a specific classical
observable and a choice for a classical time. The quantization
after imposing the constraint provides the physical meaning of the
wave functions of the universe. It is supported by an analog of
the correspondence principle of nonrelativistic quantum mechanics
that relates a quantum system to its classical counterpart.
Finally in Section~8, we offer a survey of our main results and
present our concluding remarks. The Appendix includes the
calculations that are useful but not of primary interest.

Throughout this article, we identify the separable Hilbert space
defined by an inner product with the Cauchy completion of the
corresponding inner product space and view the operators acting in
this Hilbert space as densely defined \cite{reed-simon}.
Furthermore, we use the terms `self-adjoint' and `Hermitian'
interchangeably. One must account for the difference whenever the
domain of the corresponding operator is a proper subset of the
Hilbert space \cite{reed-simon}. This leads to a similar type of
technical complications that are already present in
nonrelativistic quantum mechanics and can be dealt with similarly
\cite{fulling}.\footnote{See Ref.~\cite{kretschmer-szymanowski}
for a straightforward treatment of the issue of domains in the
context of the theory of pseudo-Hermitian operators that is used
in Ref.~\cite{cqg} and the present work.}

\section{Hilbert Space Problem for Klein-Gordon-Type Fields}
\label{s2}

In order to construct a quantum mechanics of a Klein-Gordon-type
field, one needs to promote the vector space
     \be
     {\cal H}:=\left\{\psi:\R\to\tilde{\cal H}~|~
     \ddot\psi(t)+D\psi(t)=0\right\}
     \label{eq-star-H}
     \end{equation}
of solutions of the field equation~(\ref{kgt}) to a Hilbert space,
i.e., construct a positive-definite inner product on ${\cal H}$.
One way of doing this is to construct an inner product
$\cbr\cdot,\cdot\ckt$ on ${\cal H}$ using the inner product
$\br\cdot|\cdot\kt$ on $\tilde{\cal H}$, i.e., define the inner
product $\cbr\psi,\phi\ckt$ of a pair of fields $\psi,\phi\in{\cal
H}$ in terms of their values $\psi(t)$ and $\phi(t)$ that belong
to $\tilde{\cal H}$.

It is very easy to choose an expression involving $\psi(t)$ and
$\phi(t)$ that satisfies the defining axioms of an inner product.
More difficult is to make sure that such an expression yields a
well-defined function $\cbr\cdot,\cdot\ckt:{\cal H}^2\to\C$, i.e.,
a pair $\psi,\phi\in{\cal H}$ determines $\cbr\psi,\phi\ckt$
uniquely. For example consider setting
    \be
    \cbr\psi,\phi\ckt=\br\psi(t)|\phi(t)\kt + \lambda^2
    \br\dot\psi(t)|\dot\phi(t)\kt
    \label{example}
    \end{equation}
where $\br\cdot|\cdot\kt$ is the inner product of $\tilde{\cal H}$
and $\lambda\in\R-\{0\}$ is a constant having the dimension of
$t$. This expression satisfies all the requirements of an inner
product for every choice of $t$.\footnote{If one sets $\lambda=0$
in (\ref{example}), this equation violates the condition that if
the norm of a vector vanishes then that vector must be the zero
vector. This is because there are nonzero solutions of~(\ref{kgt})
that vanish at a given $t$. This problem does not arise for
$\lambda\neq 0$, because for a nonzero Klein-Gordon-type field the
value of the field and its $t$-derivative cannot simultaneously
vanish.} Yet it fails to yield a well-defined inner product on
${\cal H}$, for one can check by differentiating the right-hand
side of (\ref{example}) that it depends on $t$. The above method
of constructing inner products on ${\cal H}$ will be effective, if
despite the explicit appearance of the variable $t$ in the
expression for $\cbr\psi,\phi\ckt$ the latter does not depend on
$t$, i.e., one obtains the same value for any choice of $t$.

In Ref.~\cite{cqg}, we employ the above method of using the
Hilbert space structure of $\tilde{\cal H}$ to construct a class
of inner products on ${\cal H}$ for the cases that the value of
$D$ at $t_0$ is a positive-definite operator.\footnote{A linear
operator is said to be positive-definite if it is a Hermitian
(self-adjoint) operator with a positive spectrum. In particular,
it is an invertible operator.} In this article we shall again
suppose that $D$ satisfies this condition. We will comment on the
Klein-Gordon-type equations that violate this condition in
Section~3.

If the operator $D$ does not depend on $t$, we will call the field
equation (\ref{kgt}) (respectively the field $\psi$) {\em
stationary}. A typical example is the free Klein-Gordon equation
(\ref{kg}).

The unique solution of a Klein-Gordon-type equation (\ref{kgt})
fulfilling the initial conditions
    \be
    \psi(t_0)=\psi_0,~~~~~~~~~~~~~~~\dot\psi(t_0)=\dot\psi_0,
    \label{ini-condi}
    \end{equation}
is
    \be
    \psi(t)=C(t,t_0)\psi_0+S(t,t_0)\dot\psi_0,
    \label{stat}
    \end{equation}
where $C(t,t_0)$ and $S(t,t_0)$ are a pair of linear operators
acting in $\tilde{\cal H}$ and satisfying
    \bea
    \ddot C(t,t_0)+D\,C(t,t_0)&=&0,~~~~~~~~~~C(t_0,t_0)=1,~~~~~~~
    \dot C(t_0,t_0)=0,
    \label{C=}\\
    \ddot S(t,t_0)+D\,S(t,t_0)&=&0,~~~~~~~~~~S(t_0,t_0)=0,~~~~~~~
    \dot S(t_0,t_0)=1.
    \label{S=}
    \eea
Here $1$ stands for the identity operator of $\tilde{\cal H}$. If
the Klein-Gordon-type equation is stationary, we can easily solve
(\ref{C=}) and (\ref{S=}) and obtain
    \bea
    C(t,t_0)&=&\cos[(t-t_0)D^{1/2}]
        :=\sum_{\ell=0}^\infty
        \frac{(-1)^\ell(t-t_0)^{2\ell}}{(2\ell)!}\,
        D^\ell,
    \label{C=2}\\
    S(t,t_0)&=&\sin[(t-t_0)D^{1/2}]D^{-1/2}
        :=\sum_{\ell=0}^\infty
        \frac{(-1)^\ell(t-t_0)^{2\ell+1}}{(2\ell+1)!}\,D^\ell.
    \label{S=2}
    \eea

For a stationary Klein-Gordon-type field with a positive-definite
$D$, the positive-definite inner products constructed in
\cite{cqg} have the form
    \bea
    \cbr\psi_1,\psi_2\ckt &=&
    \frac{1}{2}\left[\br\psi_1(t)|L_+\psi_2(t)\kt+
    \br\dot\psi_1(t)|L_+D^{-1}\dot\psi_2(t)\kt+\right.\nn\\
    &&\left.
    i(\br\psi_1(t)|L_-D^{-1/2}\dot\psi_2(t)\kt-
    \br\dot\psi_1(t)|L_-D^{-1/2}\psi_2(t)\kt)\right],
    \label{inv-inn-prod}
    \eea
where $\psi_1,\psi_2\in{\cal H}$, $\br\cdot|\cdot\kt$ is the inner
product of $\tilde{\cal H}$, and $L_\pm$ are Hermitian operators
acting in $\tilde{\cal H}$ such that $A_\pm:=L_+\pm L_-$ are
positive-definite operators commuting with $D$. For a
nonstationary Klein-Gordon-type field, one can construct a class
of inner products on ${\cal H}$ that depend on the choice of
$t_0$, i.e., an initial value of $t$. They have the form
    \be
    \cbr\psi_1,\psi_2\ckt_{t_0}:=
    \left.\cbr\psi_1,\psi_2\ckt\right|_{t=t_0},
    \label{time-dep}
    \end{equation}
where the expression on the right-hand side is obtained by
evaluating (\ref{inv-inn-prod}) at $t_0$.

The construction of the inner products (\ref{inv-inn-prod}) is a
direct consequence of the following two basic principles.
    \begin{itemize}
    \item[](I) As a vector space, the Hilbert space ${\cal H}$ has a
    dual interpretation, namely as the space of solutions $\psi$ of
    the field equation (\ref{kgt}) and as the space of all possible
    initial data $(\psi_0,\dot\psi_0)$ for this equation.
    \item[](II) In order to identify ${\cal H}$ with the Hilbert
    space of state vectors for a quantum system (admitting a
    probability interpretation), the inner product that turns
    ${\cal H}$ into a Hilbert space must be positive-definite.
    \end{itemize}

(I) follows from the linearity of the field equation (\ref{kgt})
which implies that as a vector space ${\cal H}$ is isomorphic to
the space $\tilde{\cal H}^2$ of the initial data $(\psi_0,
\dot\psi_0)$ or equivalently the space $\tilde{\cal H}\otimes\C^2$
of the two-component state vectors
    \be
    \Psi_0:=\Psi(t_0),
    \label{initial}
    \end{equation}
where $\Psi$ is the two-component field
    \be
    \Psi:=\left(\begin{array}{c}
    \psi+i\lambda\dot\psi\\
        \psi-i\lambda\dot\psi\end{array}\right),
    \label{2-com}
    \end{equation}
$\lambda\in\R-\{0\}$ is an arbitrary parameter\footnote{$\lambda$
belongs to the subgroup $GL(1,\R)$ of a $GL(2,\C)$ symmetry group
of the two-component formulation of the Klein-Gordon-type field
equations, \cite{jpa-98}.} having the dimension of $t$,
$\psi\in{\cal H}$, and $\dot\psi:\R\to\tilde{\cal H}$ is defined
by $\dot\psi(t)= \frac{d}{dt}\psi(t)$.

The identification of ${\cal H}$ with $\tilde{\cal H}\otimes\C^2$
leads to the idea of employing a two-component Schr\"odinger
formulation of the Klein-Gordon-type fields, \cite{foldy-1956,FV}.
Using the two-component  fields (\ref{2-com}) and the Hamiltonian
    \be
    H=\frac{\hbar}{2}\,\left(\begin{array}{cc}
    \lambda D+\lambda^{-1}& \lambda D-\lambda^{-1}\\
    -\lambda D+\lambda^{-1}& -\lambda D-\lambda^{-1}
    \end{array}\right),
    \label{2-com-H}
    \end{equation}
we can express the field equation (\ref{kgt}) as the Schr\"odinger
equation
    \be
    i\hbar\dot\Psi(t)=H\Psi(t).
    \label{sch-eq}
    \end{equation}

The values $\Psi(t)$ of the two-component field $\Psi$ belong to
$\tilde{\cal H}\otimes\C^2$. If we endow the latter with the inner
product $\br\cdot,\cdot\kt$ defined by
    \be
    \br\xi,\zeta\kt=\sum_{a=1}^2 \br\xi^a|\zeta^a\kt,
    \label{inner-2}
    \end{equation}
for all
    \[\xi=:\left(\begin{array}{c}\xi^1\\
    \xi^2\end{array}\right), \zeta=:\left(\begin{array}{c}\zeta^1\\
    \zeta^2\end{array}\right)\in \tilde{\cal H}\otimes\C^2,\]
and denote the corresponding Hilbert space, namely $\tilde{\cal
H}\oplus\tilde{\cal H}$, by ${\cal H}_\star$, we can view
$\Psi(t)$ as elements of ${\cal H}_\star$ and identify $H$ with a
linear operator acting in ${\cal H}_\star$. In this article, we
will confine our attention to the two-component state vectors that
belong to ${\cal H}_\star$ (have finite $\br\cdot,\cdot\kt$-norm)
and will define the adjoint $T^\dagger$ of any linear operator $T$
acting in $\tilde{\cal H}\otimes\C^2$ using the inner product
(\ref{inner-2}). Following this convention, we can easily show
that the Hamiltonian (\ref{2-com-H}) satisfies
    \be
    H^\dagger=\sigma_3 H\sigma_3,
    \label{H-sHs}
    \end{equation}
where $\sigma_3$ is the diagonal Pauli matrix (See (\ref{pauli})
below.) Noting that $\sigma_3^{-1}=\sigma_3$ and recalling that a
linear operator $H$ is said to be pseudo-Hermitian if there is an
invertible, Hermitian, linear operator $\eta$ such that
$H^\dagger=\eta H\eta^{-1}$, we see that the Hamiltonian
(\ref{2-com-H}) is pseudo-Hermitian.

Moreover, using the assumption that $D$ is a positive-definite
operator, we can easily check that $H$ is diagonalizable and has a
real spectrum. These observations together with (II) above suggest
the use of the characterization theorems of
Refs.~\cite{p1,p2,p4,p7} to construct positive-definite inner
products $\bbr\cdot|\cdot\kkt_{\tilde\eta}$ on $\tilde{\cal
H}\otimes\C^2$ that are invariant under the time-evolution
generated by the Hamiltonian $H$. The equivalence of $\tilde{\cal
H}\otimes\C^2$ with ${\cal H}$ then leads one to use the inner
products $\bbr\cdot|\cdot\kkt_{\tilde\eta}$ to construct the inner
products (\ref{inv-inn-prod}) and subsequently (\ref{time-dep}).
In the remainder of this section we sketch the derivation of
(\ref{inv-inn-prod}) as given in Ref.~\cite{cqg}. This allows us
to fix the notation and introduce the necessary tools that we will
use in the rest of the article.

First, we recall that as any inner product on $\tilde{\cal
H}\otimes\C^2$, $\bbr\cdot|\cdot\kkt_{\tilde\eta}$ may be
expressed in the form \cite{kato}:
    \be
    \bbr\xi|\zeta \kkt_{\tilde\eta}=
    \br\xi,\tilde\eta\,\zeta\kt,
    \label{tilde}
    \end{equation}
where $\xi,\zeta\in\tilde{\cal H}\otimes\C^2$ and $\tilde\eta$ is
a positive-definite operator acting in $\tilde{\cal
H}\otimes\C^2$.

Consider a stationary Klein-Gordon-type field, where $D$ and $H$
are $t$-independent. Then supposing that the operator $\tilde\eta$
and consequently the inner product~(\ref{tilde}) do not depend on
$t$, one can show \cite{p1,cqg} that the requirement of the
invariance ($t$-independence) of
$\bbr\Psi(t)|\Phi(t)\kkt_{\tilde\eta}$, for any two solutions
$\Psi(t),\Phi(t)$ of the Schr\"odinger equation (\ref{sch-eq}), is
equivalent to the condition:
    \be
    H^\dagger=\tilde\eta H\tilde\eta^{-1},
    \label{ph}
    \end{equation}
i.e., $H$ is $\tilde\eta$-pseudo-Hermitian \cite{p1}. According to
Refs.~\cite{p4,p8}, any positive-definite operator $\tilde\eta$
fulfilling (\ref{ph}) may be expressed in the form
    \be
    \tilde\eta=\sum_\nu |\tilde\Phi_\nu\kt\br\tilde\Phi_\nu|,
    \label{t-eta}
    \end{equation}
where $|\tilde\Phi_\nu\kt$ form a complete set of eigenvectors of
$H^\dagger$, $\nu$ is a spectral label\footnote{The summation over
$\nu$ means a summation over the discrete spectrum and an
integration over the continuous spectrum.}, and for all
$\xi,\zeta\in \tilde{\cal H}\otimes\C^2$, $|\xi\kt\br\zeta|$ acts
on two-component vectors $\chi\in\tilde{\cal H}\otimes\C^2$
according to
    \be
    |\xi\kt\br\zeta|\chi=\br\zeta,\chi\kt\,\xi.
    \label{project}
    \end{equation}

The eigenvalue problem for $H$ and $H^\dagger$ is easily solved
\cite{jpa-98,cqg}. The common eigenvalues of $H$ and $H^\dagger$
are
    \be
    E_{\pm,n}=\pm\,\hbar\,\omega_n,
    \label{eg-va}
    \end{equation}
where $\omega_n$ are the positive square root of the eigenvalues
$\omega_n^2\in\R^+$ of $D$. A set of eigenvectors of $H$ and
$H^\dagger$ are respectively given by
    \be
    \Psi_{\pm,n}=\left(\begin{array}{c}
    \lambda^{-1}\pm\omega_n\\
    \lambda^{-1}\mp\omega_n\end{array}\right)\phi_n,~~~~~~~~
    \Phi_{\pm,n}=\frac{1}{4}\,\left(\begin{array}{c}
    \lambda\pm\omega_n^{-1}\\
    \lambda\mp\omega_n^{-1}\end{array}\right)\phi_n,
    \label{eg-ve}
    \end{equation}
where $\phi_n$ form a complete set of orthonormal eigenvectors of
$D$, i.e.,
    \be
    D\phi_n=\omega_n^2\phi_n,~~~~~~~~~
    \br\phi_{n'}|\phi_n\kt=\delta_{n',n},~~~~~~~~~
    \sum_n|\phi_n\kt\br\phi_n|=1.
    \label{eg-va-D}
    \end{equation}
One can check that $\Psi_{\pm,n}$ and $\Phi_{\pm,n}$ form a
complete biorthonormal system of eigenvectors of $H$ and
$H^\dagger$, i.e., they satisfy
    \bea
    &&H\Psi_{\epsilon,n}=E_{\epsilon,n}\Psi_{\epsilon,n},~~~~~~
    H^\dagger\Phi_{\epsilon,n}=E_{\epsilon,n}\Phi_{\epsilon,n},
    \label{eg-va-H}\\
    && \br\Phi_{\epsilon,n},\Psi_{\epsilon',n'}\kt=
    \delta_{\epsilon,\epsilon'}\delta_{n,n'},~~~~~~~~~~
    \sum_\epsilon\sum_n |\Psi_{\epsilon,n}\kt\br\Phi_{\epsilon,n}|
    =1,
    \label{bior}
    \eea
where $\epsilon,\epsilon'=\pm$ and $n,n'$ are arbitrary spectral
(and degeneracy) labels for $D$.\footnote{The eigenvalues
$\omega_n^2$ of $D$ may be degenerate. Here we suppress the
degeneracy labels and allow for $\omega_n^2$ with different $n$ to
coincide.}

Substituting the eigenvectors $\Phi_{\pm,n}$ of (\ref{eg-ve}) for
$\tilde\Phi_\nu$ in (\ref{t-eta}), we find a particular
positive-definite operator $\tilde\eta$ that satisfies (\ref{ph}).
Using the spectral resolution of $D$, we can show that this
operator has the from
    \be
    \eta_+=\frac{1}{8}\,\left(\begin{array}{cc}
    \lambda^2+D^{-1}&\lambda^2-D^{-1}\\
    \lambda^2-D^{-1}&\lambda^2+D^{-1}\end{array}\right).
    \label{eta=}
    \end{equation}
Furthermore, because any other complete set of eigenvectors of
$H^\dagger$ may be obtained from $\Phi_{\pm,n}$ through the action
of an invertible operator $B$ that commutes with $H^\dagger$, we
can express the most general positive-definite operator
$\tilde\eta$ satisfying (\ref{ph}) as
    \be
    \tilde\eta=A^\dagger \eta_+ A,
    \label{tilde-eta=}
    \end{equation}
where $A=B^\dagger$ is an invertible linear operator commuting
with $H$, \cite{cqg,p7}. The latter condition implies that
    \be
    A=\sum_\epsilon\sum_n a_{\epsilon,n}
    |\Psi_{\epsilon,n}\kt\br\Phi_{\epsilon,n}|,
    \label{A=}
    \end{equation}
for some nonzero complex numbers $a_{\epsilon,n}$.

In view of (\ref{eg-ve}), we can express the operator $A$ in terms
of $D$ and its eigenvectors $\phi_n$ directly. This results in
    \be
    A=A_++A_-A',
    \label{A=D}
    \end{equation}
where
    \bea
    A_\pm&:=&\frac{1}{2}\,\sum_n(a_{+,n}\pm a_{-,n})
    |\phi_n\kt\br\phi_n|\, \sigma_0,
    \label{A+-}\\
    A'&:=&\frac{1}{2}\, \left[(\sigma_3+i\sigma_2)\lambda D^{1/2}+
        (\sigma_3-i\sigma_2)\lambda^{-1} D^{-1/2}\right],
    \label{A-zero}
    \eea
$\sigma_0$ is the $2\times 2$ unit matrix, and
    \be
    \sigma_1=\left(\begin{array}{cc}
    0 & 1 \\
    1 & 0\end{array}\right),~~~~~~~
    \sigma_1=\left(\begin{array}{cc}
    0 & -i \\
    i & 0\end{array}\right),~~~~~~~
    \sigma_3=\left(\begin{array}{cc}
    1 & 0 \\
    0 & -1\end{array}\right),
    \label{pauli}
    \end{equation}
are the Pauli matrices. Now, substituting (\ref{A=D}) and
(\ref{eta=}) in (\ref{tilde-eta=}), and carrying out the necessary
calculations, we find
    \be
    \tilde\eta=\frac{1}{8}\,\left(\begin{array}{cc}
    L_+(\lambda^2+D^{-1})+2\lambda L_-D^{-1/2} &
    L_+(\lambda^2-D^{-1})\\
    L_+(\lambda^2-D^{-1}) &
    L_+(\lambda^2+D^{-1})-2\lambda L_-D^{-1/2} \end{array}\right),
    \label{tilde-eta2}
    \end{equation}
where
    \be
    L_\pm:=\frac{1}{2}\sum_n (|a_{+,n}|^2\pm |a_{-,n}|^2)
    |\phi_n\kt\br\phi_n|.
    \label{L=}
    \end{equation}
Finally, introducing
    \be
    \cbr\psi_1,\psi_2\ckt:=
    \lambda^{-2}\bbr\Psi_1(t)|\Psi_2(t)\kkt_{\tilde\eta},
    \label{inn=inn}
    \end{equation}
in which $\Psi_i$ are related to $\psi_i$ according to
(\ref{2-com}), and using (\ref{tilde}), we arrive at the
expression~(\ref{inv-inn-prod}).

One can use the field equation (\ref{kgt}) to check that the
$t$-derivative of the right-hand side of (\ref{inv-inn-prod})
vanishes identically. Therefore, $\cbr\cdot,\cdot\ckt$ is a
well-defined inner product on ${\cal H}$. It can be evaluated at
any value of $t$, e.g., at $t_0$:
    \be
    \cbr\psi_1,\psi_2\ckt=\lambda^{-2}
    \bbr\Psi_1(t_0)|\Psi_2(t_0)\kkt_{\tilde\eta}.
    \label{inn=inn2}
    \end{equation}

For a nonstationary Klein-Gordon-type field, $D$ and $H$ depend on
$t$. In this case the right-hand side of (\ref{inv-inn-prod})
fails to be $t$-independent, and this relation does not provide a
well-defined inner product on ${\cal H}$. But there is a
well-defined positive-definite inner product on ${\cal H}$ which
reduces to (\ref{inv-inn-prod}) for the cases where $D$ is
$t$-independent \cite{cqg}. This inner product turns out to be
given by a direct generalization of (\ref{inn=inn2}), namely
(\ref{time-dep}). It is well-defined provided that a choice of
$t_0$ is made. We will elaborate on the general positive-definite
inner products on the solution space of nonstationary
Klein-Gordon-type fields in Section~3.

Each choice of a positive-definite inner product on ${\cal H}$
determines a quantum mechanics of the corresponding
Klein-Gordon-type field. The quantum observables are the linear
operators $o:{\cal H}\to{\cal H}$ that are Hermitian with respect
to the chosen inner product. A quantum system associated with a
Klein-Gordon-type field is uniquely determined by the choice of
the Hilbert space (a positive-definite inner product on) ${\cal H}$
and an observable called the Hamiltonian. The results reported in
Ref.~\cite{cqg} solve the Hilbert space problem for a
Klein-Gordon-type field. But they do
not explain how one should fix a particular inner product on
${\cal H}$. Neither do they provide an explicit construction of
the observables or describe the form and meaning of the possible
Hamiltonians. The present article aims at addressing these issues.
The first step in this direction is to determine the most general
positive-definite inner product on ${\cal H}$.

\section{General Form of an Inner Product on ${\cal H}$}

In standard nonrelativistic canonical quantum mechanics, the
Hilbert space is (up to its dimension if it is finite-dimensional)
unique. This follows from the fact that any two separable Hilbert
spaces (with the same dimension if they are finite-dimensional)
are related by a unitary transformation \cite{reed-simon}. Yet in
order to specify a physical system one must fix the Hilbert space
${\cal H}$ and specify a Hamiltonian operator $H:{\cal H}\to{\cal
H}$. A pair $({\cal H},H)$ determines a quantum system uniquely,
but a quantum system may be described by different (actually
infinitely many pairs) $({\cal H},H)$.

Let ${\cal H}_i$, with $i\in\{1,2\}$, be Hilbert spaces with inner
products $\br\cdot,\cdot\kt_i$, $H_i:{\cal H}_i\to{\cal H}_i$ be
Hermitian operators, and $S_i$ be the quantum system determined by
the pair $({\cal H}_i,H_i)$. Suppose that there exists a
time-independent unitary linear transformation\footnote{This means
\cite{reed-simon} that for all $\psi_1,\phi_1\in{\cal H}$, $ \br
U\psi_1,U\Phi_1\kt_2=\br\psi_1|\phi_1\kt_1$, alternatively for all
$\psi_1\in{\cal H}_1$ and $\psi_2\in{\cal H}_2$, $\br
U\psi_1,\psi_2\kt_2=\br\psi_1,U^{-1}\psi_2\kt_1$, i.e.,
$U^\dagger=U^{-1}$.} $U:{\cal H}_1\to{\cal H}_2$ such that $H_2=U
H_1 U^{-1}$. Then $S_1$ and $S_2$ are physically equivalent, i.e.,
there is a one-to-one correspondence between the states and
observables of $S_1$ and $S_2$ such that all the physically
measurable quantities, namely the transition amplitude between the
states and the expectation values of the observables, are
invariant under this correspondence. Clearly, the observables
$O_2:{\cal H}_2\to{\cal H}_2$ of $S_2$ are related to the
observables $O_1:{\cal H}_1\to{\cal H}_1$ of $S_1$ via the unitary
similarity transformation $O_2=U O_1 U^{-1}$.

Next, suppose that a quantum system $S_2$ is determined by a
Hilbert space ${\cal H}_2$ and a Hamiltonian $H_2$. Let ${\cal
H}_1$ be a vector space which is isomorphic to ${\cal H}_2$, i.e.,
there is an invertible linear transformation $U:{\cal H}_1\to{\cal
H}_2$. Then one can use $U$ to induce a Hilbert space structure on
${\cal H}_1$ and define a Hermitian operator $H_1:=U^{-1}H_2 U$
such that the quantum system $S_1$ corresponding to the pair
$({\cal H}_1,H_1)$ is physically equivalent to $S_2$. The inner
product $\br\cdot,\cdot\kt_1:{\cal H}_1^2\to\C$ induced by $U$ is
defined by
    \[\br\psi_1,\phi_1\kt_1:=\br U\psi_1,U\phi_1\kt_2,\]
where $\psi_1,\phi_1$ are arbitrary elements of ${\cal H}_1$ and
$\br\cdot,\cdot\kt_2$ is the inner product of ${\cal H}_2$. By
construction $U$ is a unitary transformation and the quantum
system $S_1$ is physically equivalent to $S_2$. The observables
$O_1$ of $S_1$ are clearly related to the observables $O_2$ of
$S_2$ via $O_1=U^{-1}O_2 U$.

As we shall demonstrate below, the construction of the inner
products~(\ref{time-dep}) given in \cite{cqg} and reviewed in the
preceding section is an example of the application of the above
method of inducing an inner product on a vector space from an
isomorphic Hilbert space.

Consider an arbitrary positive-definite inner product
$\pbr\cdot,\cdot\pkt$ on the vector space ${\cal H}$ of
(\ref{eq-star-H}), and let for all $t\in\R$, $U_t:{\cal
H}\to\tilde{\cal H}\otimes\C^2$ be defined by
        \be
        U_t\psi:=\lambda^{-1}\Psi(t),
        \label{U-sub-t}
        \end{equation}
where $\psi\in{\cal H}$ is arbitrary and $\Psi(t)$ is the
two-component field (\ref{2-com}) evaluated at $t$. It is not
difficult to see that $U_t$ is an invertible linear map. If
$\ppbr\cdot,\cdot\ppkt$ denotes the inner product on $\tilde{\cal
H}\otimes\C^2$ induced by $U_t^{-1}$, then for all
$\psi_1,\psi_2\in{\cal H}$
        \be
        \ppbr U_t\psi_1,U_t\psi_2\ppkt=\pbr\psi_1,\psi_2\pkt.
        \label{Thm-1}
        \end{equation}
In view of (\ref{U-sub-t}), we can write (\ref{Thm-1}) in the form
        \be
        \ppbr\Psi_1(t),\Psi_2(t)\ppkt=
        \lambda^2\pbr\psi_1,\psi_2\pkt,
        \label{thm-2}
        \end{equation}
where $\Psi_i$ is the two-component field (\ref{2-com}) associated
with $\psi_i$ for $i\in\{1,2\}$. By construction, $\Psi_i(t)$ are
solutions of the Schr\"odinger equation~(\ref{sch-eq}).
Furthermore, because the right-hand side of (\ref{thm-2}) does not
depend on $t$, the inner product $\ppbr\cdot,\cdot\ppkt$ is
invariant under the dynamics generated by the Hamiltonian
(\ref{2-com-H}).

As we mentioned earlier, being an inner product on $\tilde{\cal
H}\otimes\C^2$, $\ppbr\cdot,\cdot\ppkt$ has the form
        \be
        \ppbr~\cdot~,~\cdot~\ppkt=
        \bbr~\cdot~|~\cdot~\kkt_{\eta}=
        \br~\cdot~,\eta~\cdot~\kt,
        \label{thm-3}
        \end{equation}
for some positive-definite operator $\eta$ acting in
$\tilde{\cal H}\otimes\C^2$. Next we enforce the condition that
the dynamics generated by the Hamiltonian $H$ leaves the inner
product (\ref{thm-3}) of any two evolving two-component state
vectors $\Psi_i(t)$ invariant. Considering the possibility that
$\eta$ may depend on $t$, we then find
    \be
    \br\Psi_1(t),\eta(t)\Psi_2(t)\kt=
    \br\Psi_1(t'),\eta(t')\Psi_2(t')\kt,~~~~~~~~~~~~\forall
    t,t'\in\R.
    \label{t=t}
    \end{equation}
In terms of the evolution operator
    \be
    U(t,t_0)={\cal T}\:e^{-\frac{i}{\hbar}\int_{t_0}^t H(t')dt'},
    \label{U(t,t_0)}
    \end{equation}
for the Hamiltonian (\ref{2-com-H}), where ${\cal T}$ is the
time-ordering operator, (\ref{t=t}) takes the form
    \be
    U(t,t')^\dagger\eta(t) U(t,t')=\eta(t'),
    ~~~~~~~~~~~~\forall t,t'\in\R.
    \label{u-e-u=e}
    \end{equation}
%\footnote{By definition \cite{p8}, an
%invertible linear operator ${\cal X}$ acting in a Hilbert space is
%said to be $\eta$-pseudo-unitary for a Hermitian, invertible,
%linear operator $\eta$ acting in the same space, if ${\cal
%X}^{-1}=\eta^{-1} {\cal X}^\dagger\eta$ or alternatively ${\cal
%X}^\dagger\eta {\cal X}=\eta$.}
If we differentiate both sides of this equation with respect
to $t$, we find, using the Schr\"odinger equation
    \be
    i\hbar \frac{\partial}{\partial t}\,U(t,t')=H\; U(t,t'),
    \label{sch-eq-u}
    \end{equation}
that $\eta$ satisfies\footnote{Equation~(\ref{dyn-inv}) is the
defining relation for a dynamical invariant
\cite{lewis-riesenfeld} of a non-Hermitian Hamiltonian. This is to
be expected as the invariance of the inner product (\ref{thm-3})
means that the matrix elements of $\eta$ in an evolving basis of
the Hilbert space are constant, \cite{nova}.}
    \be
    i\hbar\; \dot{\eta}= H^\dagger \eta- \eta\, H.
    \label{dyn-inv}
    \end{equation}
It is not difficult to check that the general solution of this
equation is given by fixing $t'$ and $\eta(t')$ in
(\ref{u-e-u=e}), i.e., we have \cite{cqg}
    \be
    \eta(t)=U(t,t_0)^{-1\dagger}\eta_0 U(t,t_0)^{-1},
    \label{thm-4.1}
    \end{equation}
where $t_0$ is an initial value of $t$ and $\eta_0$ is a
$t$-independent positive-definite operator acting in $\tilde{\cal
H}\otimes\C^2$.

Substituting~(\ref{thm-4.1}) in (\ref{thm-3}) and using
(\ref{thm-2}), we have
     \be
     \pbr\psi_1,\psi_2\pkt=\lambda^{-2}
     \bbr\Psi_1(t_0)|\Psi_2(t_0)\kkt_{\eta_0}.
     \label{tilde-inv-2}
     \end{equation}
Therefore, every positive-definite inner product on ${\cal H}$ is
determined by a $t_0$ and a $t$-independent positive-definite
operator $\eta_0$ that yields a solution (\ref{thm-4.1}) of
(\ref{dyn-inv}). This equation has a constant ($t$-independent)
solution if and only if $H$ is $\eta$-pseudo-Hermitian with
respect to a $t$-independent positive-definite operator $\eta$.
This is the case for the stationary Klein-Gordon-type fields where
$H$ is $t$-independent and $\eta=\tilde\eta$ has the general form
(\ref{tilde-eta2}). In view of (\ref{inn=inn}) and (\ref{thm-2}),
we see that in this case the inner product $\pbr\cdot,\cdot\pkt$
coincides with the inner product $\cbr\cdot,\cdot\ckt$ given by
(\ref{inv-inn-prod}).

If we set $t=t_0$ in (\ref{thm-4.1}) we find that
$\eta(t_0)=\eta_0$. This provides yet another characterization of
the general positive-definite inner products $\pbr\cdot,\cdot\pkt$
on ${\cal H}$, namely that the latter are given by the functions
${\cal F}$ mapping $\R$ into the set of all positive-definite
operators acting in $\tilde{\cal H}\otimes\C^2$. Every such
function provides an assignment of a $t$-independent
positive-definite operator $\eta_0$ to each $t_0\in\R$ according
to $\eta_0={\cal F}(t_0)$.

The inner products (\ref{inv-inn-prod}) and (\ref{time-dep})
correspond to a function ${\cal F}_0$ that assigns to each $t_0$
the value $\tilde\eta_0$ of the operator $\tilde\eta$ of
(\ref{inn=inn}) at $t_0$, i.e., set $\eta_0=\tilde\eta_0$,
so that
    \be
    \pbr\psi_1,\psi_2\pkt=\lambda^{-2}
    \bbr\Psi_1(t_0)|\Psi_2(t_0)\kkt_{\tilde\eta_0}=
    \cbr\psi_1,\psi_2\ckt_{t_0}.
    \label{tilde-inv-2-prime}
    \end{equation}
Clearly, ${\cal F}_0$ is not the most general choice for ${\cal
F}$. It is however a distinguished choice, for it leads to a
constant solution of (\ref{dyn-inv}) and consequently defines a
$t$-independent inner product on $\tilde{\cal H}\otimes\C^2$ for
the stationary Klein-Gordon-type fields.

In the remainder of this article we shall only be concerned with
the inner products (\ref{time-dep}) that reduce to
(\ref{inv-inn-prod}) for stationary Klein-Gordon-type fields. We
can justify this restriction by noting that every other inner
product leads to the same Hilbert space structure on ${\cal H}$,
i.e., the corresponding Hilbert spaces are related by unitary
maps. The latter are  actually easy to construct.

Let $\pbr\cdot,\cdot\pkt$ be the inner product associated with an
arbitrary choice for the positive-definite operator $\eta_0$ and
$\cbr\cdot,\cdot\ckt_{t_0}$ be an inner product of the form
(\ref{time-dep}). Then
    \bea
    \pbr\psi_1,\psi_2\pkt&=&
    \lambda^{-2}\bbr\Psi_1(t_0)|\Psi_2(t_0)\kkt_{\eta_0}=
    \lambda^{-2}\br\Psi_1(t_0),\eta_0\Psi_2(t_0)\kt,
    \label{thm-20}\\
    \cbr\psi_1,\psi_2\ckt_{t_0}&=&
    \lambda^{-2}\bbr\Psi_1(t_0)|\Psi_2(t_0)\kkt_{{\tilde\eta}_0}=
    \lambda^{-2}\br\Psi_1(t_0), {\tilde\eta}_0\Psi_2(t_0)\kt.
    \label{thm-21}
    \eea
where $\tilde\eta_0=\tilde\eta(t_0)$ is a $t$-independent
positive-definite operators acting in $\tilde{\cal
H}\otimes\C^2$ such that $H(t_0)$ is
$\tilde\eta_0$-pseudo-Hermitian.  Because $\eta_0$
(respectively $\tilde\eta_0$) is a positive-definite operator, it
has a positive-definite square root $\rho_0$ (respectively
$\tilde\rho_0$). Clearly, ${\cal U}:=\tilde\rho_0^{-1}\rho_0$ is
an invertible operator acting in $\tilde{\cal H}\otimes\C^2$,
    \[{\cal U}^\dagger\tilde\eta_0{\cal U}=
    \rho_0\tilde\rho_0^{-1}\tilde\rho_0^2
    \tilde\rho_0^{-1}\rho_0=\eta_0,\]
and for all $\xi,\zeta\in\tilde{\cal H}\otimes\C^2$
    \be
    \bbr\xi|\zeta\kkt_{\eta_0}=\br\xi,\eta_0\zeta\kt=
    \br\xi,{\cal U}^\dagger \tilde\eta_0{\cal U}\zeta\kt=
    \br{\cal U}\xi,\tilde\eta_0{\cal U}\zeta\kt=
    \bbr{\cal U}\xi|{\cal U}\zeta\kkt_{\tilde\eta_0}.
    \label{x-z=}
    \end{equation}
Next, consider the invertible operator $U_1:=U_{t_0}$, i.e.,
    \be
    U_1\psi:=\lambda^{-1}\Psi_0,
    \label{U-1}
    \end{equation}
where $U_t$ is defined in (\ref{U-sub-t}) and $\Psi_0$ is the
initial two-component state vector~(\ref{initial}), and let ${\cal
U}':=U_1^{-1}{\cal U} U_1$. Then Because both  ${\cal
U}:\tilde{\cal H}\otimes\C^2\to\tilde{\cal H}\otimes\C^2$ and
$U_1:{\cal H}\to\tilde{\cal H}\otimes\C^2$ are invertible, ${\cal
U}'$ is an invertible linear operator acting in ${\cal H}$.
Furthermore, in view of  (\ref{thm-20}) --
(\ref{U-1}), we have for all $\psi_1,\psi_2\in{\cal H}$
    \[\pbr\psi_1,\psi_2\pkt=
    \bbr U_1\psi_1(t_0)|U_1\psi_1\kkt_{\eta_0}
    =\bbr {\cal U}U_1\psi_1(t_0)|
    {\cal U}U_1\psi_1\kkt_{\tilde\eta_0}
    =\cbr{\cal U}'\psi_1,{\cal U}'\psi_2\ckt_{t_0}.\]
This shows that ${\cal U}'$ is a unitary operator relating the
inner products (\ref{thm-20}) and (\ref{thm-21}). Therefore, as
far as the physical content of a quantum mechanics of a
Klein-Gordon-type field is concerned, these inner products are
equivalent, and we can suppose, without loss of generality, that
the operator $\eta_0$ appearing in (\ref{tilde-inv-2}) is such
that $H(t_0)$ is $\eta_0$-pseudo-Hermitian, i.e.,
$\eta_0=\tilde\eta_0$.

The above discussion of the most general positive-definite inner
product on ${\cal H}$ may be extended with minor revisions to the
cases that the operator $D$ appearing in the field
equation~(\ref{kgt}) is Hermitian but not positive-definite. If
$D$ is still invertible, i.e., its spectrum does not include zero,
then we can pursue using the inner product (\ref{time-dep})
provided that we let $D_0$ stand for the value of $\sqrt{D^2}$ at
$t_0$ and substitute $D_0$ for $D$ in (\ref{inv-inn-prod}) and
consequently (\ref{time-dep}). Clearly, in this case $D^2$ is a
positive-definite operator possessing a unique positive-definite
square root $\sqrt{D^2}$. We can also use the argument given in
the preceding paragraph to relate any other inner product on
${\cal H}$ to the inner product (\ref{time-dep}) constructed in
this way. If $D$ is a general Hermitian operator with a nontrivial
null space, then we can still use the inner product
(\ref{time-dep}) provided that we replace the operator $D$
appearing in (\ref{inv-inn-prod}) by a positive-definite operator
$D'$. Again all the choices for $D'$ would lead to
unitarily equivalent Hilbert spaces.

As we noted in Section~2, in this paper we consider the cases
where $D$ is a positive-definite operator and study, without loss
of generality, the consequences of endowing ${\cal H}$ with the
inner product (\ref{time-dep}).

\section{Quantum Mechanics of Stationary Klein-Gordon-Type Fields}
\label{s3}

In this section we study the quantum mechanics of stationary
Klein-Gordon-type fields, i.e., suppose that $D$ does not depend
on $t$. First, we introduce the following notation:
\begin{itemize}
\item[~] $q_{L_\pm}=$ the quantum mechanics defined by the Hilbert
space ${\cal H}$ having (\ref{inv-inn-prod}) as its inner product;
\item[~] ${\cal H}_{L_\pm}=$ the Hilbert space obtained by
endowing $\tilde{\cal H}\otimes\C^2$ with the inner product given
by (\ref{tilde}) where $\tilde\eta$ has the form
(\ref{tilde-eta2}); \item[~] $Q_{L_\pm}=$ the quantum mechanics
defined by the Hilbert space ${\cal H}_{L_\pm}$; \item[~]
$S_{L_\pm}=$ the quantum system determined by the Hilbert space
${\cal H}_{L_\pm}$ and the Hamiltonian (\ref{2-com-H}).
\end{itemize}

\subsection{Equivalence of  $q_{L_{\pm}}$ and $Q_{L_{\pm}}$}

In Section~3, we used the operator $U_t$ of (\ref{U-sub-t}) to
relate the one- and two-component Klein-Gordon-type fields. This
operator is clearly $t$-dependent. For stationary
Klein-Gordon-type fields one can use a $t$-independent invertible
operator to relate one- and two-component fields, namely the
operator $U_1:{\cal H}\to\tilde{\cal H}\otimes\C^2$ defined in
(\ref{U-1}). As we mentioned in Section~3, $U_1$ is an invertible
linear operator. $U_1^{-1}:\tilde{\cal H}\otimes\C^2\to{\cal H}$
is the operator that maps each two-component state vector
    \[\xi=\left(\begin{array}{c}\xi^1\\ \xi^2\end{array}\right)
    \in\tilde{\cal H}\otimes\C^2\]
into the solution $\psi$ of the field equation~(\ref{kgt})
satisfying the initial conditions
    \[\psi(t_0)=\frac{1}{2}\,(\xi^1+\xi^2),~~~~~~~~~\dot\psi(t_0)=
    \frac{1}{2i}\,(\xi^1-\xi^2).\]
By virtue of (\ref{stat}), we have (for all $t$)
    \bea
    (U_1^{-1}\xi)(t)&=&\frac{1}{2}\,\left\{
    \cos[(t-t_0)D^{1/2}](\xi^1+\xi^2)-i\sin[(t-t_0)D^{1/2}]D^{-1/2}
    (\xi^1-\xi^2)\right\}\nn\\
    &=& V(t)\xi^1+V(t)^\dagger\xi^2,
    \label{U-1=}
    \eea
where
    \be
    V(t):=\frac{1}{2}\,\left\{\cos[(t-t_0)D^{1/2}]
    -i\sin[(t-t_0)D^{1/2}]D^{-1/2}\right\}.
    \label{V}
    \end{equation}
Furthermore, we can use (\ref{inn=inn2}) to show that for all
$\psi_1,\psi_2\in{\cal H}$,
    \be
    \cbr\psi_1,\psi_2\ckt=\lambda^{-2}
    \bbr\Psi_1(t_0)|\Psi_2(t_0)\kkt_{\tilde\eta}=
    \lambda^{-2}\bbr \lambda U_1\psi_1|
    \lambda U_1\psi_2\kkt_{\tilde\eta}
    =\bbr U_1\psi_1|U_1\psi_2\kkt_{\tilde\eta}.
    \label{u1}
    \end{equation}
Hence $U_1$ is a unitary operator, and $q_{L_\pm}$ and $Q_{L_\pm}$
are equivalent.

\subsection{Hamiltonians for a stationary Klein-Gordon-type field}
\label{s3-1}

Having established the unitarity of $U_1$, we can use $S_{L_\pm}$
to define a quantum system $s_{L_\pm}$ with the Hilbert space
${\cal H}$ and the Hamiltonian
    \be
    h:=U_1^{-1} H U_1.
    \label{h=}
    \end{equation}
Note that because $H$ is a Hermitian operator with respect to the
inner product (\ref{tilde}) and $U_1$ is unitary, $h$ is Hermitian
with respect to the inner product (\ref{inv-inn-prod}) on ${\cal H}$.

It is interesting to see how the Hamiltonian operator $h$ acts on
the (one-component) Klein-Gordon-type fields $\psi$. Letting
    \be
    \phi:=h\psi
    \label{eq0}
    \end{equation}
and making use of (\ref{2-com}), (\ref{U-1}), and $U_1h=HU_1$
which is equivalent to (\ref{h=}), we find\footnote{It is
remarkable that the free parameter $\lambda$ that enters the
expressions for $H$ and the initial two-component fields
associated with $\psi$ and $\phi$ disappears in the final result
of this calculation.}
    \be
    \phi(t_0)=i\hbar\dot\psi(t_0),~~~~~~~~~~
    \dot\phi(t_0)=-i\hbar D\psi(t_0).
    \label{eq}
    \end{equation}
By construction, $\phi$ is a solution of the field
equation~(\ref{kgt}) satisfying the initial conditions (\ref{eq}).
Because $D$ is $t$-independent, $\tilde\psi:=i\hbar\dot\psi$
satisfies both the field equation~(\ref{kgt}) and the initial
conditions
    \be
    \tilde\psi(t_0)=i\hbar\dot\psi(t_0),~~~~~~~~~~
    \dot{\tilde\psi}(t_0)=-i\hbar D\psi(t_0).
    \label{eq50}
    \end{equation}
Therefore, by the uniqueness of the solution (\ref{stat}) of the
initial-value problem for equation~(\ref{kgt}), we have
     \bea
    \phi(t)&=&i\hbar\dot\psi(t),
     \label{eq1}\\
    \dot\phi(t)&=&-i\hbar D\psi(t),
     \label{eq2}
     \eea
for all $t$.\footnote{The consistency of (\ref{eq1}) and
(\ref{eq2}) is equivalent to the field equation (\ref{kgt}).}
According to (\ref{eq0}) and (\ref{eq1}),
    \be
    h\psi=i\hbar\dot\psi.
    \label{h-psi}
    \end{equation}
The fact that $\dot\psi(t)$ satisfies the field equation
(\ref{kgt}), i.e., $\dot\psi\in{\cal H}$, is consistent with the
requirement that $h$ maps ${\cal H}$ into itself.

The remarkable resemblance of (\ref{h-psi}) to the time-dependent
Schr\"odinger equation of nonrelativistic quantum mechanics is
actually misleading. Unlike the latter which determines the
$t$-dependence of an evolving state vector, (\ref{h-psi})
describes the action of the operator $h$ on the space ${\cal H}$
of the solutions of the field equation (\ref{kgt}), i.e., it is
the definition of $h$. Because $h$ is obtained through a unitary
transformation from the Hamiltonian $H$, one might expect that it
should be possible to determine the $t$-dependence of the value
$\psi(t)$ of the field\footnote{It is the value of the field
$\psi$ (and note $\psi$ itself) that depends on $t$.} using a
Schr\"odinger-like equation involving $h$. In order to see that
this is actually not the case, we evaluate both sides of
(\ref{h-psi}) at $t$. This yields
    \be
    i\hbar\frac{d}{dt}\psi(t)=(h\psi)(t),
    \label{sch-eq-1com-1}
    \end{equation}
where the action of $h$ on $\psi$ is determined by (\ref{h-psi}).
Now, we compare (\ref{sch-eq-1com-1}) with a time-dependent
Schr\"odinger equation that is satisfied by $\psi(t)\in\tilde{\cal
H}$. The latter would have the form
    \be
    i\hbar\frac{d}{dt}\psi(t)=\tilde h \psi(t),
    \label{sch-eq-1com-2}
    \end{equation}
for a Hamiltonian operator $\tilde h$ acting in the Hilbert space
$\tilde{\cal H}$.

Equations~(\ref{sch-eq-1com-1}) and (\ref{sch-eq-1com-2}) are
fundamentally different. One cannot solve (\ref{sch-eq-1com-1})
for $\psi(t)$ without using the field equation~(\ref{kgt}). This
is because, by construction, the domain of the definition of $h$
is the space of solutions of~(\ref{kgt}).

In terms of $h$, (\ref{eq2}) takes the form
    \be
    h\dot\psi=-i\hbar D\psi,
    \label{eq3}
    \end{equation}
where again $\psi\in{\cal H}$ is an arbitrary solution of
(\ref{kgt}) and $D\psi$ is the element of ${\cal H}$ defined by
$(D\psi)(t):=D\psi(t)$.\footnote{Because $D$ is assumed to be
$t$-independent, for all $\psi\in{\cal H}$, $(D\psi)(t)=D\psi(t)$
also satisfies~(\ref{kgt}). Hence $D\psi\in{\cal H}$, and $D$ may
be viewed as a linear operator acting in ${\cal H}$.} Combining
(\ref{eq3}) and (\ref{h-psi}), we find $h^2\psi=\hbar^2 D\psi$.
Therefore, as operators acting in ${\cal H}$, $h^2$ and $\hbar^2
D$ coincide,
    \be
    h^2=\hbar^2 D.
    \label{eq4}
    \end{equation}

Next, we apply $\tilde h$ to both sides of (\ref{sch-eq-1com-2}).
If we now assume that $\tilde h$ is $t$-independent and $\psi(t)$
appearing in (\ref{sch-eq-1com-2}) is an arbitrary solution of
the field equation~(\ref{kgt}), we obtain
    \be
    \tilde h^2\psi(t)=\hbar^2 D\psi(t).
    \label{eq5}
    \end{equation}
It is essential to observe that this equation only holds if
$\psi(t)$ satisfies the field equation (\ref{kgt}). As an operator
equation in $\tilde{\cal H}$, $\tilde h^2=\hbar^2 D$ does not
hold. For if it did, we could use the fact that $D$ is a
positive-definite operator to infer that $\tilde h=\hbar\Delta$
where $\Delta$ is a square root of $D$. The resulting
Schr\"odinger equation (\ref{sch-eq-1com-2}), with any choice for
$\Delta$, is not equivalent to the Klein-Gordon-type
equation~(\ref{kgt}). This is because we may choose the initial
data $(\psi_0,\dot\psi_0)$ so that $\dot\psi_0\neq i\Delta
\psi_0$. In this case the Schr\"odinger equation
(\ref{sch-eq-1com-2}) with $\tilde h=\hbar\Delta$ and any square
root $\Delta$ of $D$ is violated at $t=t_0$. This argument is
a manifestation of the rather obvious fact that the $t$-dependence
of Klein-Gordon-type fields cannot be described by a time-dependent
Schr\"odinger equation defined in the Hilbert space $\tilde{\cal H}$.

The Hamiltonian $h$ does not determine the $t$-dependence of the
value $\psi(t)$ of a given field $\psi$. It generates a
time-evolution in the space ${\cal H}$ of fields. The
corresponding time-dependent Schr\"odinger equation reads
    \bea
    i\hbar\frac{d}{dt}\psi_t&=&h\psi_t,
    \label{sch-eq-field}\\
    \psi_{t_0}&=&\phi,
    \label{int-cond}
    \eea
where for each value of $t\in\R$, $\psi_t\in{\cal H}$. We will
denote by $\psi_t(t')$ the value of the field $\psi_t$ at $t'$.
Clearly $\psi_t(t')\in\tilde{\cal H}$.

Before we explore the consequences of the Schr\"odinger
equation~(\ref{sch-eq-field}), we wish to comment on the precise
meaning of the term `time' used in this article. We will identify
a real variable with a time-parameter if and only if it is the
evolution-parameter associated with a (Hermitian) Hamiltonian
operator acting in the Hilbert space ${\cal H}$. It is the choice
of a Hamiltonian operator that decides whether a real parameter is
to be qualified as a measure of time. For example the parameter
$t$ appearing in the Schr\"odinger equation~(\ref{sch-eq-field})
is a time-parameter whereas the parameter $t$ appearing in the
argument of the value $\psi(t)$ of a Klein-Gordon-type field
$\psi\in{\cal H}$ cannot be termed as `time' unless we adopt a
Hamiltonian operator acting in ${\cal H}$ that generates
$t$-translations of the fields.\footnote{This point is not
properly taken into account in the terminology used in
Ref.~\cite{cqg}. There the Klein-Gordon-type equation~(\ref{kgt})
is viewed as an evolution equation and the parameter $t$ appearing
in this equation is called time to reflect this point of view.}
{\em A priori} such a (Hermitian) Hamiltonian may or may not
exist. The above definition of time is consistent provided that
one does not apply it in a discussion of the two-component fields.
This is justified by noting that the two-component formulation of
a Klein-Gordon-type equation, in which the argument $t$ of the
value $\psi(t)$ of the fields $\psi$ is the evolution-parameter
for the Hamiltonian (\ref{2-com-H}), involves the arbitrary
unphysical parameter $\lambda$.\footnote{As pointed out in
\cite{jpa-98}, the freedom in the choice of $\lambda$ is related to
a $GL(1,\C)$ symmetry of the two-component formulation of the
Klein-Gordon-type fields. Indeed, one may consider $t$-dependent
$\lambda$'s in which case the corresponding gauge symmetry is
local. Physically it signifies a $t$-reparameterization
symmetry of the two-component formalism.}

Next, we evaluate both sides of (\ref{sch-eq-field}) at $t'$,
i.e., substitute $t'$ for $t$ and $\psi_t(t')$ for $\psi(t)$ in
(\ref{sch-eq-1com-1}). We also replace the ordinary derivatives
with partial derivatives as $t$ and $t'$ are independent
variables. Then using (\ref{h-psi}) we find
    \[\frac{\partial}{\partial t}\,\psi_t(t')=
    \frac{\partial}{\partial t'}\,\psi_t(t').\]
This together with (\ref{int-cond}) yields
    \be
    \psi_t(t')=\phi(t'+t-t_0).
    \label{t-trans}
    \end{equation}
The particularly simple mixing of the parameters $t$ and $t'$
suggests that the argument $t'$ of the value $\phi(t')$ of the
field $\phi$ has the same physical meaning as the time-parameter
$t$.

Moreover, if we write the Schr\"odinger equation
(\ref{sch-eq-field}) in the form
    \be
    \psi_t=u(t,t_0)\phi,
    \label{u}
    \end{equation}
with
    \be
    u(t,t_0):=e^{-i(\frac{t-t_0}{\hbar})h},
    \label{u=}
    \end{equation}
and let $\delta t:=t-t_0$, we can express (\ref{t-trans}) in the
form
    \be
    e^{-i(\frac{\delta t}{\hbar})\,h}\:\phi(t')=\phi(t'+\delta t).
    \label{eq6}
    \end{equation}
This equation shows that the time-evolution generated by the
Hamiltonian $h$ corresponds to $t$-translations of the stationary
Klein-Gordon-type fields. In other words, $h$ is the generator of
the $t$-translations in the space ${\cal H}$. This is another
indication that the choice of $h$ as the Hamiltonian is equivalent
to identifying the parameter $t$ appearing in the defining field
equation~(\ref{kgt}) with time.

Next, we use the expression~(\ref{h-psi}) for the Hamiltonian $h$
to determine the energy eigenstates of the system. Substituting
this expression in the eigenvalue equation
    \be
    h\psi_n=e_n\psi_n,
    \label{eg-va-eq-h}
    \end{equation}
we obtain $i\hbar\dot\psi_n=e_n\psi_n$. Consequently,
$\psi_n(t)=e^{-ie_n(t-t_0)/\hbar} \psi_n(t_0)$. Now imposing the
condition that $\psi_n$ is a solution of the field equation
(\ref{kgt}) with a $t$-independent $D$, we find that $\psi_n(t_0)$
is an eigenvector of $D$ with eigenvalue $e_n^2/\hbar^2$. Hence,
in light of (\ref{eg-va-D}), $e_n=\pm\hbar\omega_n$, and
$\psi_n(t_0)$ is an eigenvector of $D$ with eigenvalue
$\omega_n^2$. In particular, $\psi_{\pm,n}\in{\cal H}$ defined by
    \be
    \psi_{\pm,n}(t):=N_{\pm,n}\,e^{\mp i\omega_n t}\phi_n,
    \label{eg-ve-h}
    \end{equation}
form a set of energy eigenvectors, where $N_{\pm,n}\in\C-\{0\}$
are arbitrary normalization constants.

We have obtained the Hamiltonian $h$ and consequently the quantum
system $s_{L_\pm}$ by using a particular unitary operator mapping
${\cal H}$ onto ${\cal H}_{L_\pm}$, namely $U_1$. If we choose
another unitary operator $\check U_1:{\cal H}\to{\cal H}_{L_\pm}$
to induce a quantum system in $q_{L_\pm}$ from $S_{L_\pm}$, we
will obtain an equivalent quantum system to $s_{L_\pm}$. However,
if we select a quantum system $\check S_{L_\pm}$ in $Q_{L_\pm}$
that is not equivalent to $S_{L_\pm}$, i.e., a Hamiltonian $\check
H:{\cal H}_{L_\pm}\to{\cal H}_{L_\pm}$ that is not related to $H$
by a unitary similarity transformation, and use $U_1$ or $\check
U_1$ to induce a quantum system $\check s_{L_\pm}$ from $\check
S_{L_\pm}$, then obviously $s_{L_\pm}$ and $\check S_{L_\pm}$ will
not be physically equivalent. The choice of $h$ as the Hamiltonian
of a quantum system having ${\cal H}$ as its Hilbert space is by
no means unique. Different choices for the Hamiltonian define
different notions of time-evolution in ${\cal H}$.

In summary, we have shown that there is a canonical quantum system
$s_{L_\pm}$ whose Hamiltonian $h$ generates $t$-translations in
${\cal H}$ so that $t$ plays the role of time, and that this is
not the only quantum system associated with a stationary
Klein-Gordon-type field, i.e., one can consider other Hamiltonians
with other choices for a time-parameter.

\subsection{Formulating the quantum mechanics of a stationary\\
Klein-Gordon-type field using the Hilbert space ${\cal
H}_\star$}\label{s3-2}

In Section~\ref{s3-1} we constructed a canonical quantum system
$s_{L_\pm}$ for stationary Klein-Gordon-type fields that was by
construction physically equivalent to $S_{L_\pm}$. In this section
we show that the systems $S_{L_\pm}$ (and consequently
$s_{L_\pm}$) corresponding to all possible choices for the
operators $L_\pm$ are also physically equivalent. This involves
constructing various unitary operators between the corresponding
Hilbert spaces and allows for a formulation of the quantum
mechanics of a stationary Klein-Gordon-type field having ${\cal
H}_\star$ as its Hilbert space.

It is useful to introduce the notation ${\cal H}_{0}$ for the
Hilbert space ${\cal H}_{L_\pm}$ with the choice $L_+=1$ and
$L_-=0$ and $S_0$ for the quantum system $S_{L_\pm}$ corresponding
to this choice. The inner product on ${\cal H}_0$ is
$\bbr\cdot|\cdot\kkt_{\eta_+}$ where $\eta_+$ is given by
(\ref{eta=}).

Now, consider the operator $A$ of (\ref{A=}). This is an
invertible operator acting in $\tilde{\cal H}\otimes\C^2$. In view
of (\ref{tilde}) and (\ref{tilde-eta=}), it satisfies, for all
$\xi,\zeta\in\tilde{\cal H}\otimes\C^2$,
    \[ \bbr A\xi|A\zeta\kkt_{\eta_+}=\br A\xi,\eta_+A\zeta\kt=
    \br\xi,A^\dagger\eta_+ A\zeta\kt=
    \br\xi,\tilde\eta\zeta\kt=
    \bbr\xi|\zeta\kkt_{\tilde\eta}.\]
This equation shows that $A$ is a unitary operator mapping ${\cal
H}_{L_\pm}$ to ${\cal H}_0$. It also commutes with $H$. Hence, the
quantum systems $S_{L_\pm}$ with all possible choices for
$L_{\pm}$ are actually equivalent to $S_0$.

Next, we recall that the operator $\eta_+$ defining the inner
product of ${\cal H}_0$ is a positive-definite operator.
Therefore, it has a unique positive-definite square root
$\rho:=\sqrt \eta_+$ namely
    \be
    \rho=\frac{1}{4}\left(\begin{array}{cc}
    \lambda+D^{-1/2} & \lambda-D^{-1/2}\\
    \lambda-D^{-1/2} & \lambda+D^{-1/2}\end{array}\right).
    \label{rho}
    \end{equation}
One can check that $\rho$ is a Hermitian operator acting in
$\tilde{\cal H}\otimes\C^2$ and satisfying
    \be
    \rho^2=\eta_+.
    \label{eq7}
    \end{equation}
This, in particular, implies that $\rho$ is an invertible
operator. The inverse of $\rho$ has the form
     \be
    \rho^{-1}=\left(\begin{array}{cc}
    \lambda^{-1}+D^{1/2} & \lambda^{-1}-D^{1/2}\\
    \lambda^{-1}-D^{1/2} & \lambda^{-1}+D^{1/2}\end{array}\right).
    \label{rho-inv}
    \end{equation}
We can also use $\rho^{-1}$ to induce a new quantum system
$S_\star$. This is determined by the Hilbert space obtained by
endowing $\tilde{\cal H}\otimes\C^2$ with the inner product
    \be
    \bbr\xi|\Phi\kkt:=\bbr\rho^{-1}\xi|\rho^{-1}\Phi\kkt_{\eta_+}=
    \br\rho^{-1}\xi,\eta_+\rho^{-1}\Phi\kt=
     \br\xi,\rho^{-1}\eta_+\rho^{-1}\Phi\kt=\br\xi,\Phi\kt,
    \label{eq8}
    \end{equation}
i.e., the Hilbert space ${\cal H}_\star=\tilde{\cal H}\oplus
\tilde{\cal H}$, and the Hamiltonian
    \be
    H_\star=\rho H\rho^{-1}=\hbar\left(\begin{array}{cc}
    D^{1/2} & 0\\
    0 & - D^{1/2}\end{array}\right)=\hbar D^{1/2}\sigma_3,
    \label{H-star}
    \end{equation}
which is clearly Hermitian with respect to the inner product
$\br\cdot,\cdot\kt$ of ${\cal H}_\star$.

By construction $S_0$ and $S_\star$ are physically equivalent.
This in turn implies that the quantum systems $S_{L_\pm}$ (and
subsequently $s_{L_\pm}$) are physically equivalent to $S_\star$.
The Hilbert spaces ${\cal H}_{L_\pm}$ are mapped to
${\cal H}_\star$ by the unitary operator
    \be
    U_2:=\rho A,
    \label{U-2}
    \end{equation}
and the Hilbert space ${\cal H}$ with inner product
(\ref{inv-inn-prod}) is related to ${\cal H}_\star$ by the unitary
operator
    \be
    U:=U_2 U_1=\rho A U_1,
    \label{UU}
    \end{equation}
where $U_1$ is defined by (\ref{U-1}). The following chain of
unitary mappings summarizes the above constructions.
    \[{\cal H}\stackrel{U_1}{\longrightarrow}{\cal H}_{L_\pm}
    \stackrel{A}{\longrightarrow} {\cal H}_0
    \stackrel{\rho}{\longrightarrow} {\cal H}_\star.\]
It further demonstrates the equivalence of $q_{L_\pm}$ with any
possible choice for $L_\pm$ with the quantum mechanics $Q_\star$
on the Hilbert space ${\cal H}_\star$.

In fact, the operator $U$ turns out to have a relatively simple
form. In order to see this, first we use (\ref{A=}) --
(\ref{A-zero}), (\ref{rho}), and (\ref{U-2}) to establish the
following useful identity:
    \be
    U_2=\rho A={\cal A}\:\rho,
    \label{eq30}
    \end{equation}
where
    \bea
    {\cal A}&:=&\left(\begin{array}{cc}
    {\cal A}_+ & 0 \\
    0 & {\cal A}_-\end{array}\right),
    \label{eq30.1}\\
    {\cal A}_\pm&:=& A_+\pm A_-=\sum_n a_{\pm,n}|\phi_n\kt\br\phi_n|.
    \label{eq31}
    \eea
Because the coefficients $a_{\pm,n}$ do not vanish, the operators
${\cal A}_\pm$ and consequently ${\cal A}$ are invertible. Clearly
${\cal A}^{-1}$ has the form
    \be
    {\cal A}^{-1}=\left(\begin{array}{cc}
    {\cal A}_+^{-1} & 0 \\
    0 & {\cal A}_-^{-1}\end{array}\right),
    \label{eq33}
    \end{equation}
where
    \be
    {\cal A}_\pm^{-1}=\sum_n a_{+,n}^{-1}|\phi_n\kt\br\phi_n|.
    \label{eq34}
    \end{equation}
In terms of ${\cal A}_\pm$, the operators $L_\pm$ appearing in the
expression for the inner product (\ref{inv-inn-prod}) take the
form
    \be
    L_\pm =\frac{1}{2}\,\left({\cal A}_+^\dagger{\cal A}_+\pm
    {\cal A}_-^\dagger{\cal A}_-\right).
    \label{L=2}
    \end{equation}
We can use (\ref{eq30}) to obtain the following expression
describing the action of the operator $U$ on a given field
$\psi\in{\cal H}$:
    \be
    U\psi=\frac{1}{2}\left(\begin{array}{c}
    {\cal A}_+(\psi_0+iD^{-1/2}\dot\psi_0)\\
    {\cal A}_-(\psi_0-iD^{-1/2}\dot\psi_0)\end{array}\right).
    \label{U-psi}
    \end{equation}
It is remarkable that unlike $U_1$ and $U_2$, the operator $U$
does not depend on the arbitrary parameter $\lambda$. This is also
the case for the Hamiltonian $H_\star$ of (\ref{H-star}).
Therefore, similarly to $s_{L_\pm}$, the quantum system $S_\star$
is also independent of the unphysical free parameter $\lambda$.
This in turn means that we can use $S_\star$ to describe the
quantum mechanics of stationary Klein-Gordon-type fields and the
corresponding dynamics generated by the Hamiltonian $h$.

\subsection{Quantum observables for a stationary Klein-Gordon-type
field}\label{s3-3}

Having obtained the explicit form of the unitary operator $U$ and
established the $\lambda$-independence of $Q_\star$, we can
construct the observables of $q_{L_\pm}$ from those of $Q_\star$.
The latter are Hermitian operators acting in ${\cal H}_\star$.

In view of the fact that ${\cal H}_\star=\tilde{\cal
H}\oplus\tilde{\cal H}$, any Hermitian operator acting in ${\cal
H}_\star$ has the form
    \be
    O_\star=\left(\begin{array}{cc}
    \tilde O_1 & \tilde{\cal O}\\
    \tilde{\cal O}^\dagger & \tilde O_2\end{array}\right),
    \label{O-star}
    \end{equation}
where $\tilde O_1,\tilde O_2$, and $\tilde{\cal O}$ are linear
operators acting in $\tilde{\cal H}$, and $\tilde O_1,\tilde O_2$
are Hermitian.

We can express the observables $o$ of $q_{L_\pm}$ in terms of the
observables $O_\star $ of $Q_\star$ according to
    \be
    o=U^{-1} O_\star U=U_1^{-1}U_2^{-1}O_\star U_2 U_1,
    \label{O=UOU-1}
    \end{equation}
where $U$ is the unitary operator (\ref{UU}) mapping ${\cal H}$ to
${\cal H}_\star$ and $U_1$ and $U_2$ are respectively given by
(\ref{U-1}) and (\ref{U-2}).

Before deriving the explicit form of the observables $o$, we wish
to comment on the interpretation of $o$ and $O_\star$ as the
Schr\"odinger- or Heisenberg-picture observables for the quantum
systems $s_{L_\pm}$ and $S_\star$, respectively. Because $U_1$
maps a field $\psi$ to the initial two-component state vector
$\lambda^{-1}\Psi_0$, there is no difference between viewing
$O_\star$ as a Schr\"odinger- or Heisenberg-picture observable.
This follows from the observation that at the initial time $t_0$
the Schr\"odinger- and Heisenberg-picture observables coincide. If
we identify $O_\star$ with a Schr\"odinger-picture observable
$O_\star^{(S)}$ and denote the corresponding Heisenberg-picture
observable by $O_\star^{(H)}$, we have, for all $t\in\R$,
    \be
    O_\star^{(S)}=O_\star,
    ~~~~~~~~~~~
    O_\star^{(H)}(t)=U_\star(t,t_0)^{-1} O_\star U_\star(t,t_0),
    \label{S-H1}
    \end{equation}
where
    \bea
    U_\star(t,t_0)&:=&e^{-i(\frac{t-t_0}{\hbar})H_\star}=
    \left(\begin{array}{cc}
    {\cal U}(t,t_0) & 0\\
    0 & {\cal U}(t,t_0)^{-1}\end{array}\right),
    \label{t-evolv}\\
    {\cal U}(t,t_0)&:=& e^{-i(t-t_0)D^{1/2}}.
    \label{U-t}
    \eea
In this case we can use the physical equivalence of $Q_\star$ and
$q_{L_\pm}$ to identify the Schr\"odinger-picture observables
$o^{(S)}$ and Heisenberg-picture observables $o^{(H)}$ for
$q_{L_\pm}$ according to
    \be
    o^{(S)}=o,~~~~~~~~~~~~~o^{(H)}(t)=u(t,t_0)^{-1}o\; u(t,t_0),
    \label{s-h1}
    \end{equation}
where $u(t,t_0)$ is given by (\ref{u=}). Similarly, if we identify
$O_\star$ with a Heisenberg-picture observable, we find
    \bea
    O_\star^{(S)}&=&U_\star(t,t_0)\;O_\star(t) U_\star(t,t_0)^{-1},
    ~~~~~~~~~~~O_\star^{(H)}(t)=O_\star(t),
    \label{S-H2}\\
    o^{(S)}&=&u(t,t_0)\;o(t)\; u(t,t_0)^{-1},
    ~~~~~~~~~~~o^{(H)}(t)=o(t).
    \label{s-h2}
    \eea

Next, we obtain the explicit form of the observables $o$ of
$q_{L_\pm}$ by computing their action on an arbitrary
$\psi\in{\cal H}$. In order to do so, first we employ
(\ref{O=UOU-1}), (\ref{O-star}), (\ref{U-1}), (\ref{UU}),
(\ref{U-2}), (\ref{eq33}), (\ref{rho}), (\ref{rho-inv}), and
(\ref{ini-condi}) to determine the initial conditions for $o\psi$.
After a lengthy calculation we find
    \bea
    &&(o\psi)(t_0)=(J_++K_+)\psi_0+i(J_-+K_-)D^{-1/2}\dot\psi_0,
    \label{o-psi-ini-1}\\
    &&\left.\frac{d}{dt}[(o\psi)(t)]\right|_{t=t_0}=D^{1/2}
    \left[i(-J_++K_+)\psi_0+(J_--K_-)D^{-1/2}\dot\psi_0\right],
    \label{o-psi-ini-2}
    \eea
where
    \be
    J_\pm:=\frac{1}{2}\left({\cal A}_+^{-1}\tilde O_1{\cal A}_+
            \pm{\cal A}_+^{-1}\tilde{\cal O}{\cal A}_-\right),
    ~~~~~~~~~~~
    K_\pm:=\frac{1}{2}\left({\cal A}_-^{-1}
            \tilde{\cal O}^\dagger{\cal A}_+\pm
            {\cal A}_-^{-1}\tilde{O}_2{\cal A}_-\right).
    \label{J-K=}
    \end{equation}
Using (\ref{U-1=}), we then obtain, for all $t$,
    \be
    (o\psi)(t)=
    \left[{\cal U}(t,t_0) J_++
    {\cal U}(t,t_0)^\dagger K_+\right]\psi_0+
    i\left[{\cal U}(t,t_0) J_-+{\cal U}(t,t_0)^\dagger K_-
    \right]D^{-1/2}\dot\psi_0,
    \label{o-psi}
    \end{equation}
where ${\cal U}(t,t_0)$ is given by (\ref{U-t}).
Equation~(\ref{o-psi}) provides the general form of the
observables of $q_{L_\pm}$.

As a consistency check of our analysis, we compute the observable
$o$ of $q_{L_\pm}$ associated with the Hamiltonian $H_\star$ of
$S_\star$. Setting $O_\star=H_\star$ in (\ref{O-star}), we find
$\tilde O_1=-\tilde O_2=\hbar D^{1/2}$ and $\tilde{\cal O}=0$.
These in turn imply $J_\pm=\mp K_\pm=\hbar D^{1/2}/2$.
Substituting these relations and (\ref{U-t}) in (\ref{o-psi}) and
doing the necessary algebra, we find the expected result: $o=h$.

Next, we calculate the matrix element $\cbr\psi,o\phi\ckt$
associated with a pair $\psi,\phi$ of elements of ${\cal H}$. In
view of the fact that $U$ is a unitary operator and using
(\ref{U-psi}), we have
    \bea
    \cbr\psi,o\phi\ckt=\br U\psi,O_\star U\phi\kt&=&
    \br\psi_0|({\cal J}_++{\cal K}_+)\phi_0\kt+
    \br\dot\psi_0|D^{-1/2}({\cal J}_--{\cal
    K}_-)D^{-1/2}\dot\phi_0\kt+\nn\\
    &&i[\br\psi_0|({\cal J}_-+{\cal K}_-)D^{-1/2}\dot\phi_0\kt+
    \br\dot\psi_0|D^{-1/2}(-{\cal J}_++{\cal K}_+)\phi_0\kt],~~~
    \label{matrix-element}
    \eea
where $(\psi_0,\dot\psi_0)$ and $(\phi_0,\dot\phi_0)$ are
respectively the initial date for the fields $\psi$ and $\phi$,
and
    \be
    {\cal J}_\pm:=\frac{1}{2}\,\left(
        {\cal A}_+^\dagger\tilde O_1{\cal A}_+
            \pm{\cal A}_+^\dagger\tilde{\cal O}{\cal A}_-\right),~~~~~~
    {\cal K}_\pm:=\frac{1}{2}\left(
        {\cal A}_-^\dagger\tilde{\cal O}^\dagger{\cal A}_+
            \pm{\cal A}_-^\dagger\tilde{O}_2{\cal A}_-\right).
    \label{cur-J-K=}
    \end{equation}
It is not difficult to check that
$\cbr\psi,o\phi\ckt^*=\cbr\phi,o\psi\ckt$. This confirms the fact
that $o$ is Hermitian with respect to the inner product
$\cbr\cdot,\cdot\ckt$.

In the Appendix, we explore the form of the observables of
$q_{L_\pm}$ for a particular two-parameter family of $L_\pm$'s
that is of importance in the study of Klein-Gordon fields in a
Minkowski spacetime, \cite{cqg}.

\section{Quantum Mechanics of Nonstationary Klein-Gordon-Type
Fields}\label{s4}

Consider the case where the operator $D$ depends on $t$. Then
again the solutions $\psi$ of the field equation (\ref{kgt}) are
given by (\ref{stat}). However, the operators $C(t,t_0)$ and
$S(t,t_0)$ no longer satisfy (\ref{C=2}) and (\ref{S=2}). In fact, a
closed-form formula for $C(t,t_0)$ and $S(t,t_0)$ is not known.
They may be expressed as infinite series involving certain
time-ordered products of $D$.

As we discussed in Section~3, we can adopt, without loss of
generality, the positive-definite inner product (\ref{time-dep})
on ${\cal H}$. We can express it more explicitly as
    \be
    \cbr\psi,\phi\ckt_{t_0}=
    \frac{1}{2}\left[\br\psi_0|L_{0+}\phi_0\kt+
    \br\dot\psi_0|L_{0+}D_0^{-1}\dot\phi_0\kt+
    i(\br\psi_0|L_{0-}D_0^{-1/2}\dot\phi_0\kt-
    \br\dot\psi_0|L_{0-}D_0^{-1/2}\psi_0\kt)\right],
    \label{inv-inn-prod-zero}
    \end{equation}
where $\psi,\phi\in{\cal H}$, $D_0:=D(t_0)$, and $L_{0\pm}$ are
arbitrary Hermitian operators such that $A_{0\pm}:=L_{0+}\pm
L_{0-}$ are positive-definite operators commuting with $D_0$.
Clearly, $L_{0\pm}=L_\pm(t_0)$ and $A_{0\pm}= A_\pm(t_0)$, where
$L_\pm$ and $A_\pm$ are respectively given by (\ref{L=}) and
(\ref{A+-}).

In Section~3, we showed that the inner
product~(\ref{inv-inn-prod-zero}) was obtained from an invariant
inner product $\bbr\cdot|\cdot\kkt_{\eta(t)}$ on $\tilde{\cal
H}\otimes\C^2$ where
    \[\eta(t):=U(t,t_0)^{-1\dagger}\tilde\eta_0
    U(t,t_0)^{-1},\]
$\tilde\eta_0=\tilde\eta(t_0)$ and $\tilde\eta$ is given
by~(\ref{tilde-eta2}). In this section we attempt to extend the
analysis of Section~4 to the nonstationary Klein-Gordon-type
fields. Again, we begin our presentation by introducing some
useful notation:
\begin{itemize}
\item[~] ${\cal H}=$ the Hilbert space obtained by endowing the
    space of solutions of the field equation~(\ref{kgt}) with the
    inner product~(\ref{inv-inn-prod-zero});
\item[~] $q_{L_{0\pm}}=$ the quantum mechanics determined by
    the Hilbert space ${\cal H}$;
\item[~] ${\cal H}_{L_{0\pm}}=$ the Hilbert space obtained
    by endowing $\tilde{\cal H}\otimes\C^2$ with the
    inner product~$\bbr\cdot|\cdot\kkt_{\eta(t)}$;
\item[~] $Q_{L_{0\pm}}=$ the quantum mechanics determined by
    the Hilbert space ${\cal H}_{L_{0\pm}}$;
\item[~] $S_{L_{0\pm}}=$ the quantum system determined by
    the Hilbert space ${\cal H}_{L_{0\pm}}$ and the
    Hamiltonian $H$ of (\ref{2-com-H}).
\end{itemize}

We also wish to recall that, by definition \cite{p8}, an
invertible linear operator ${\cal X}$ acting in a Hilbert space is
said to be $\eta'$-pseudo-unitary for a Hermitian, invertible,
linear operator $\eta'$ acting in the same space, if ${\cal
X}^{-1}=\eta^{'-1} {\cal X}^\dagger\eta'$ or alternatively ${\cal
X}^\dagger\eta' {\cal X}=\eta'$.

\subsection{Equivalence of  $q_{L_{0\pm}}$ and $Q_{L_{0\pm}}$}

Let $\tilde\eta_0=\tilde\eta(t_0)$ where $\tilde\eta$ is given by
(\ref{tilde-eta=}), i.e., $\tilde\eta_0:\tilde{\cal H}
\otimes\C^2\to\tilde{\cal H}\otimes\C^2$ is the most general
positive-definite operator such that $H(t_0)$ is
$\tilde\eta_0$-pseudo-Hermitian. Suppose that ${\cal
V}:\tilde{\cal H} \otimes\C^2\to\tilde{\cal H}\otimes\C^2$ is an
arbitrary (possibly time-dependent) $\tilde\eta_0$-pseudo-unitary
operator, so that
    \be
    {\cal V}^\dagger\tilde\eta_0 {\cal V}=\tilde\eta_0,
    \label{p-unitary}
    \end{equation}
and $U'_1:{\cal H}\to{\cal H}_{L_{0\pm}}$ is defined by
    \be
    U'_1:=U(t,t_0){\cal V}\,U_1,
    \label{U-1-prime}
    \end{equation}
where $U(t,t_0)$ is the evolution operator (\ref{U(t,t_0)}) for
the Hamiltonian $H$. Then because $U(t,t_0)$, ${\cal V}$, and
$U_1$ are invertible linear operators, so is $U_1'$. Furthermore,
a straightforward calculation shows that, for all
$\psi,\phi\in{\cal H}$,
    \bea
    \bbr U_1'\psi|U_1'\phi\kkt_{\eta(t)}&=&\lambda^{-2}\bbr
    U(t,t_0){\cal V}\Psi_0|U(t,t_0){\cal V}\Phi_0\kkt_{\eta(t)}
    =\lambda^{-2}\br\Psi_0|{\cal V}^\dagger\tilde\eta_0
    {\cal V}\Phi_0\kt\nn\\
    &=&
    \lambda^{-2}\br\Psi_0|\tilde\eta_0\Phi_0\kt=
     \lambda^{-2}\bbr\Psi_0|\Phi_0\kkt_{\tilde\eta_0}=
    \cbr\psi,\phi\ckt_{t_0},
    \label{eq60}
    \eea
where we have made use of (\ref{U-1-prime}), (\ref{U-1}), and
(\ref{tilde-inv-2-prime}). As seen from (\ref{eq60}), $U_1'$ is a
unitary operator manifesting the equivalence of $q_{L_{0\pm}}$ and
$Q_{L_{0\pm}}$.

\subsection{Hamiltonians for a nonstationary Klein-Gordon-type
field}

We can use $U_1'$ to induce a Hamiltonian operator $h':{\cal
H}\to{\cal H}$ from the Hamiltonian $H$ of $S_{L_{0\pm}}$.
However note that unlike the operator $U_1$, $U'_1$ is generally
$t$-dependent. This implies that the requirement that the dynamics
generated by $h'$ in ${\cal H}$ is mapped to the dynamics
generated by $H$ in ${\cal H}_{L_{0\pm}}$ is equivalent to the
condition~\cite{nova}:
    \[h'={U_1'}^{-1}H\,U'_1 -i\hbar{U_1'}^{-1}\dot U'_1.\]
Substituting (\ref{U-1-prime}) in this equation, we find
    \be
    h'=U_1^{-1} V\, U_1,
    \label{h-prime}
    \end{equation}
where
    \be
    V:=-i\hbar {\cal V}^{-1}\dot{\cal V}.
    \label{V=}
    \end{equation}
We shall denote  the quantum system associated with the Hilbert
space ${\cal H}$ and the Hamiltonian $h'$ by $s'_{L_{0\pm}}$.

As seen from the above construction, the fact that $U_1'$ is
unitary follows from the condition that {${\cal V}$ is
$\tilde\eta_0$-pseudo-unitary. It is not difficult to observe that
the converse is also true. Therefore all the unitary operators
mapping ${\cal H}$ to ${\cal H}_{L_{0\pm}}$ have the form
(\ref{U-1-prime}) for some $\tilde\eta_0$-pseudo-unitary operator
${\cal V}$. The latter form a pseudo-unitary group
$G_{\tilde\eta_0}$, \cite{p8}. In fact, because $\tilde\eta_0$ is
a positive-definite operator, this group is isomorphic to the
unitary group $U({\cal H}_\star)$ of all the unitary operators
acting in the Hilbert space ${\cal H}_\star$, \cite{p8}. The
operator ${\cal V}$, viewed as a function of $t$, traces a path in
the group $G_{\tilde\eta_0}$. If we choose ${\cal V}$ to be a
constant function, i.e., ${\cal V}$ is time-independent, $h'=0$.%
\footnote{This is similar to the quantum mechanical analog
\cite{nova} of the dynamical canonical transformation used in the
Hamilton-Jacobi formulation of classical mechanics.} It is also
interesting to note that because ${\cal V}$ is
$\tilde\eta_0$-pseudo-unitary (${\cal V}\in G_{\tilde\eta_0}$), $V$
is $\tilde\eta_0$-pseudo-Hermitian \cite{p8}. In other words,
similarly to $H(t_0)$, the operator $V$ belongs to the Lie algebra
${\cal G}_{\tilde\eta_0}$ of the group $G_{\tilde\eta_0}$.

The unitary operator $U'_1$ and consequently the Hamiltonian $h'$
are determined by the operators ${\cal V}$ belonging to the group
$G_{\tilde\eta_0}$. For a stationary Klein-Gordon-type field,
where $D=D_0$ and $\tilde\eta=\tilde\eta_0$, the Hamiltonian $H$
of $S_{L_\pm}$ is $\tilde\eta_0$-pseudo-Hermitian \cite{p1}. This
is sufficient to deduce that the evolution operator $U(t,t_0)$ is
$\tilde\eta_0$-pseudo-unitary \cite{p8}, i.e., $U(t,t_0)\in
G_{\tilde\eta_0}$. As a result, $U(t,t_0)^{-1}\in
G_{\tilde\eta_0}$, and we can take
    \be
    {\cal V}=U(t,t_0)^{-1}.
    \label{V=U}
    \end{equation}
This yields $U'_1=U_1$, $V=H$, and $h'=h$. Clearly (\ref{V=U}) is
not the only choice for ${\cal V}$. However, it is this choice
that identifies the Hamiltonian $h'$ with the generator $h$ of the
$t$-translations in ${\cal H}$ and consequently makes $t$ the
time-parameter for a stationary Klein-Gordon-type field. This
observation leads to the natural question if there is a choice for
${\cal V}$ that makes $h'$ the generator of $t$-translations in
${\cal H}$ (equivalently identifies $t$ with time) for a
nonstationary Klein-Gordon-type field. We will next show that the
answer to this question is negative.

Consider the case that $D$ does depend on $t$ and let $h:{\cal
H}\to{\cal H}$ be the operator defined by (\ref{h=}). Then
substituting (\ref{U-1}) and (\ref{2-com-H}) in (\ref{h=}), we
find that the action of $h$ on a given field $\psi\in{\cal H}$ is
described by the expressions (\ref{eq}) for the initial data of
$\phi:=h\psi$. Note however that, unlike for a stationary
Klein-Gordon-type field, the operator $D$ appearing in these
relations depends on $t$. Therefore, acting $h$ on a field $\psi$
yields a one-parameter family of fields $\phi_t$ parameterized by
$t$.\footnote{This is a manifestation of the fact that being
obtained via a $t$-independent unitary transformation from a
$t$-dependent operator (namely $H$), $h$ is $t$-dependent.}

Next, we explore the time-evolution generated by the operator $h$.
If we denote by $\check u(t,t_0)$ the time-evolution operator
associated with $h$, namely
    \be
    \check u(t,t_0):={\cal T}\, e^{-\frac{i}{\hbar}\int_{t_0}^t h(t')dt'},
    \label{time-evolve-h}
    \end{equation}
we can use (\ref{h=}) to show that
    \be
    U_1\check u(t,t_0)=U(t,t_0) U_1.
    \label{UU=UU}
    \end{equation}
Applying both sides of this equation on an initial field
$\psi_{t_0}\in{\cal H}$, denoting the evolving Klein-Gordon-type
field by $\psi_t:=\check u(t,t_0)\psi_{t_0}$, and using
(\ref{U-1}), we find
    \be
    \lambda U_1\psi_t=\Psi_t(t_0)=\Psi_{t_0}(t),
    \label{eq51}
    \end{equation}
where for all $t_1,t_2\in\R$, $\Psi_{t_1}(t_2)$ stands for the
value of the two-component field associated with $\psi_{t_1}$ at
$t_2$. In view of (\ref{2-com}) and (\ref{eq51}),
    \be
    \psi_t(t_0)=\psi_{t_0}(t),~~~~~~~~~~~~~~~~~
    \dot\psi_t(t_0):=\left.\frac{\partial}{\partial t'}
    \psi_t(t')\right|_{t'=t_0}=\dot\psi_{t_0}(t).
    \label{eq52}
    \end{equation}
Furthermore, we can check that the field $\psi'_{t_0}$ defined by
$\psi'_{t_0}(t'):=\psi_{t_0}(t'+t-t_0)$ also satisfies the initial
conditions (\ref{eq52}). Hence by the uniqueness of the solution
of the initial-value problem for (\ref{kgt}), we have
$\psi_t(t')=\psi_{t_0}(t'+t-t_0)$ for all $t'$. This equation
shows that the time-evolution generated by $h$ is a
$t$-translation of the fields; the choice of $h$ as the
Hamiltonian is equivalent to identifying $t$ with time. However
note that a time-translation does not correspond to a unitary
time-evolution in ${\cal H}$, unless for all $t$ the Hamiltonian
$H$ happens to be $\tilde\eta_0$-pseudo-Hermitian. If this
condition holds, we can choose (\ref{V=U}) and obtain $h'=h$. As
we mentioned above, this is the case for a stationary
Klein-Gordon-type field. In general this condition is not
fulfilled, $h$ is not a Hermitian operator acting in ${\cal H}$,
and $t$-translations are not unitary operators in this space.

This argument is valid even if we choose an arbitrary
positive-definite inner product on ${\cal H}$ that is not
necessarily of the form (\ref{time-dep}). Suppose that $h$ is
Hermitian with respect to some inner product $\pbr\cdot,\cdot\pkt$
on ${\cal H}$. As we have shown in Section~3, such an inner
product is obtained from an invariant inner product
$\bbr\cdot,\cdot\kkt_\eta$ on $\tilde{\cal H}\otimes\C^2$ that is
defined by a general $t$-independent positive-definite operator
$\eta_0$ according to (\ref{thm-3}) and (\ref{thm-4.1}). In light
of the fact that the operator $U_t=U(t,t_0)U_1$ defined by
(\ref{U-sub-t}) is a unitary operator mapping ${\cal H}$ equipped
with the inner product $\pbr\cdot,\cdot\pkt$ to $\tilde{\cal
H}\otimes\C^2$ with inner product $\bbr\cdot,\cdot\kkt_\eta$, the
operator
    \be
    H_t:=U_t\,h\,U_t^{-1}=U(t,t_0)U_1\,h\,U_1^{-1}U(t,t_0)^{-1}=
    U(t,t_0)\,H(t)\, U(t,t_0)^{-1},
    \label{zeq-1}
    \end{equation}
must be Hermitian with respect to $\bbr\cdot,\cdot\kkt_\eta$. This
is equivalent to saying that $H_t$ is $\eta$-pseudo-Hermitian:
$H_t^\dagger=\eta\, H_t\eta^{-1}$. Substituting (\ref{thm-4.1})
and (\ref{zeq-1}) in this equation then yields
$H(t)^\dagger=\eta_0 H(t)\,\eta_0^{-1}$. Hence, demanding that $h$
be Hermitian with respect to some inner product on ${\cal H}$
implies that for all $t$, $H(t)$ is
$\eta_0$-pseudo-Hermitian for some $t$-independent positive-definite
operator $\eta_0$.\footnote{The converse of this statement is
clearly true.} The necessary and
sufficient condition for the latter is that $H(t_0)$ be
$\eta_0$-pseudo-Hermitian and the eigenvectors of $H(t)$ do not
depend on $t$. In view of (\ref{eg-va}) -- (\ref{eg-va-D}), this
holds if and only if $D$ is $t$-independent. Therefore, for a
nonstationary Klein-Gordon-type field, one cannot identify $t$
with a time-parameter associated with the dynamics generated by a
Hermitian Hamiltonian.\footnote{As $h$ is a diagonalizable
operator with a real spectrum, one may appeal to the results of
\cite{p2} to argue for the existence of an inner product with
respect to which $h$ is Hermitian. This inner product will
necessarily be $t$-dependent and hence ill-defined. Moreover,
because of the $t$-dependence of such an inner product, the
Hermiticity of $h$ is not sufficient for the unitarity of the
dynamics it generates \cite{cqg}. One can also attempt to
construct an inner product with respect to which the dynamics
generated by $h$ is unitary, but this inner product will also be
necessarily $t$-dependent and consequently ill-defined.}

As we mentioned above, it is a $t$-dependent element ${\cal V}$
of the group $G_{\tilde\eta_0}$ that determines the Hamiltonian
$h'$ and consequently the quantum system $s'_{L_{0\pm}}$.  There
are infinitely many choices for ${\cal V}$. Among them are certain
choices that make $h'$ $t$-independent. For example, letting
    \be
    {\cal V}={\cal V}_0:=e^{i\frac{(t-t_0)}{\hbar}\,H(t_0)}
    \label{choice}
    \end{equation}
yields $h'=h'_0:=\hbar U_1H(t_0)U_1^{-1}=h(t_0)$, where $H(t_0)$
and $h(t_0)$ are respectively the Hamiltonians (\ref{2-com-H}) and
(\ref{h=}) evaluated at $t=t_0$. Clearly for stationary
Klein-Gordon-type fields, (\ref{choice}) coincides
with (\ref{V=U}).

We can use the analysis of Section~4.2 to solve the eigenvalue
problem for the Hamiltonian $h'_0$. The eigenvalues have the form
$\pm\hbar\omega_{0n}$ where $\omega_{0n}$ are the positive square
roots of the eigenvalues of $D_0$. A complete set of eigenvectors
$\psi_{\pm,n}$ of $h'$ (with eigenvalues $\pm\hbar\,\omega_{0n}$)
are given by the following initial conditions
    \[\psi_{\pm,n}(t_0)=N_{\pm,n}\phi_{0n},~~~~~~~~~~~~~
    \dot\psi_{\pm,n}(t_0)=
    \mp i N_{\pm,n}\omega_{0n}\phi_{0n}.\]
Here $N_{\pm,n}\in\C$ are normalization constants and $\phi_{0n}$
are linearly independent eigenvectors of $D_0$ corresponding to
the eigenvalue $\omega_{0n}^2$.

We conclude this section by emphasizing that for both stationary
and nonstationary Klein-Gordon-type fields the choice of a
Hamiltonian acting in ${\cal H}$ is not unique. Choosing
nonequivalent Hermitian operators as Hamiltonians acting in ${\cal
H}_{L_{0\pm}}$ and using the unitary operator $U_1'$ with any
choice of ${\cal V}\in G_{\tilde\eta_0}$ to induce a corresponding
Hamiltonian acting in ${\cal H}$ yield nonequivalent notions of
time-evolution in ${\cal H}$.

\subsection{Formulating the quantum mechanics of a nonstationary
Klein-Gordon-type field using the Hilbert space ${\cal H}_\star$}

Let $A_0,{\cal A}_0,\eta_{0+},\rho_0:\tilde{\cal
H}\otimes\C^2\to\tilde{\cal H}\otimes\C^2$, be the values of the
operators (\ref{A=D}), (\ref{eq30.1}), (\ref{eta=}), and
(\ref{rho}) at $t=t_0$, respectively, i.e.,
    \be
    A_0:=A(t_0),
    ~~~~~~~~~~~~~{\cal A}_0:={\cal A}(t_0),
    ~~~~~~~~~~~~~\eta_{0+}=\eta_+(t_0),
    ~~~~~~~~~~~~~\rho_0:=\rho(t_0),
    \label{eq54}
    \end{equation}
so that
    \bea
    \tilde\eta_0&=&A_0^\dagger\eta_{0+}A_0,
    \label{eq81a}\\
    \eta_{0+}&=&\rho_0^2,
    \label{eq81b}
    \eea
Suppose that ${\cal W},U_2':\tilde{\cal
H}\otimes\C^2\to\tilde{\cal H}\otimes\C^2$ are linear operators
such that ${\cal W}$ is $\tilde\eta_0$-pseudo-unitary (${\cal
W}\in G_{\tilde\eta_0}$):
    \be
    {\cal W}^\dagger\tilde\eta_0{\cal W}=\tilde\eta_0,
    \label{eq82}
    \end{equation}
and $U_2'$ is defined by
    \be
    U_2':=\rho_0A_0\,{\cal W}\,U(t,t_0)^{-1}.
    \label{U2-prime}
    \end{equation}
Then a simple calculation shows that, for all
$\xi,\zeta\in\tilde{\cal H}\otimes\C^2$,
    \bea
    \bbr \xi|\zeta\kkt_{\eta(t)}&=&
    \br\xi\:,\:U(t,t_0)^{-1\dagger}\tilde\eta_0
    U(t,t_0)^{-1}\zeta\kt\nn\\
    &=&\br \xi\:,\:U(t,t_0)^{-1\dagger}{\cal W}^\dagger
    \tilde\eta_0{\cal W} U(t,t_0)^{-1}\zeta\kt\nn\\
    &=&\br \xi\:,\:U(t,t_0)^{-1\dagger}{\cal W}^\dagger
    A_0^\dagger\rho_0^2A_0{\cal W} U(t,t_0)^{-1}\zeta\kt\nn\\
    &=&\br \xi\:,\: U_2^{'\dagger} U_2^{'}\zeta\kt\nn\\
    &=&\br U_2'\xi,U_2'\zeta\kt.\nn
    \eea
Therefore, $U_2'$ is a unitary operator mapping the Hilbert space
${\cal H}_{L_{0\pm}}$ to the Hilbert space ${\cal H}_\star$, and
the quantum mechanics $Q_{L_{0\pm}}$ and subsequently
$q_{L_{0\pm}}$ with all possible choices for $L_{0\pm}$ are
equivalent to $Q_\star$. The following diagram shows the unitary
mappings relating ${\cal H}, {\cal H}_{L_{0\pm}}$, and ${\cal
H}_\star$.
    \[{\cal H}\stackrel{U_1'}{\longrightarrow}{\cal H}_{L_{0\pm}}
    \stackrel{U_2'}{\longrightarrow}{\cal H}_\star.\]

We can also use $U_2^{'-1}$ to induce a Hamiltonian $H_\star'$ on
${\cal H}_\star$ from the Hamiltonian $H$ on ${\cal
H}_{L_{0\pm}}$. Again because $U_2'$ depends on $t$,
$H_\star':=U_2'H\,U_2^{'-1}+i\hbar\dot U_2'U_2^{'-1}$.
Substituting (\ref{U2-prime}) in this equation, we find
    \be
    H_\star'=\rho_0A_0 W\,A_0^{-1}\rho_0^{-1}=
    \rho_0W_+\rho_0^{-1},
    \label{H-star-prime}
    \end{equation}
where
    \be
    W:=i\hbar \dot{\cal W}\,{\cal W}^{-1},~~~~~~~~~~
    W_+:=A_0WA_0^{-1}.
    \label{W=}
    \end{equation}
The $\tilde\eta_0$-pseudo-unitarity of ${\cal W}$ implies that $W$
is an $\tilde\eta_0$-pseudo-Hermitian operator, i.e.,
    \be
    W^\dagger=\tilde\eta_0\, W\tilde\eta_0^{-1}.
    \label{ph-W}
    \end{equation}

Because $U_2':{\cal H}_{L_{0\pm}}\to{\cal H}_\star$ and $H:{\cal
H}_{L_{0\pm}}\to{\cal H}_{L_{0\pm}}$ are respectively unitary and
Hermitian operators, $H'_\star$ is a Hermitian operator acting in
the Hilbert space ${\cal H}_\star$. We can also compute
$H_\star^{'\dagger}$ directly. To do this, first we use
(\ref{ph-W}) and(\ref{eq81a}) to show that $W_+$ is
$\eta_{0+}$-pseudo-Hermitian, i.e., $W_+^\dagger=\eta_{0+}
W_+\,\eta_{0+}^{-1}$. In view of this relation and (\ref{eq81b})
and (\ref{H-star-prime}), we then have
    \[H_\star^{'\dagger}=(\rho_0W_+\rho_0^{-1})^\dagger=
    \rho_0^{-1}W_+^\dagger\rho_0=\rho_0 W_+\rho_0^{-1}=
    H'_\star.\]
Hence $H'_\star$ is indeed a Hermitian operator acting in ${\cal
H}_\star$. As seen from (\ref{H-star-prime}) it is determined by
the choice of an $\eta_{0+}$-pseudo-Hermitian operator
$W_+\in{\cal G}_{\eta_{0+}}$.

By construction, the unitary operator $U':{\cal H}\to{\cal
H}_\star$ defined by
    \be
    U':=U_2'U_1'
    \label{U-prime}
    \end{equation}
maps the Hamiltonian $h'$ of the quantum system $s'_{L_{0\pm}}$ to
the Hamiltonian $H_\star'$. Therefore, $s'_{L_{0\pm}}$ is equivalent
to the quantum system $S'_\star$ defined by the Hilbert space
${\cal H}_\star$ and the Hamiltonian $H_\star'$.

For a stationary Klein-Gordon-type field where $D$ does not depend
on $t$ and $U(t,t_0)\in G_{\tilde\eta_0}$, we can set ${\cal
W}=U(t,t_0)$ and $W=H$. Now using (\ref{H-star}) and
(\ref{H-star-prime}) and the fact that in this case $A_0$ commutes
with $H$, we see that $H_\star'=H_\star$ and the quantum system
$S'_\star$ reduces to $S_\star$.

\subsection{Quantum observables for a nonstationary
Klein-Gordon-type field}

The construction of the observables for stationary
Klein-Gordon-type fields as reported in Section~4.4 generalizes to
nonstationary Klein-Gordon type fields. Again one uses the unitary
operator $U':{\cal H}\to{\cal H}_\star$ to express the observables
$o$ of $q_{L_{0\pm}}$ in terms of the observables $O_\star$ of
$Q_\star$ according to
    \be
    o=U^{'-1}O_\star U'.
    \label{eq90}
    \end{equation}
Here $O_\star$ has the general form~(\ref{O-star}) and $U'$ is the
unitary operator~(\ref{U-prime}).

Substituting (\ref{U-1-prime}) and (\ref{U2-prime}) in
(\ref{U-prime}), we find
    \be
    U'=\rho_0 A_0{\cal X}U_1,~~~~~~~~~~~~{\cal X}:={\cal WV}.
    \label{U-prime-2}
    \end{equation}
As seen from this equation $U'$ is determined by the choice of an
element ${\cal X}$ of the group $G_{\tilde\eta_0}$. We start our
derivation of the general form of the observables $o$ by showing
that we can absorb the arbitrariness in ${\cal X}$ in the
arbitrariness in the form of the observables $O_\star$. Using the
$\tilde\eta_0$-pseudo-unitarity of ${\cal X}$, i.e.,
    \be
    {\cal X}^\dagger=\tilde\eta_0{\cal X}^{-1}\tilde\eta_0,
    \label{pu-X}
    \end{equation}
and (\ref{eq81a}), we can establish the identity
    \be
    (A_0{\cal X})^{-1}=\tilde\eta_0^{-1}(A_0{\cal
    X})^\dagger\eta_{0+}.
    \label{eq500}
    \end{equation}
Furthermore, we introduce a linear operator $R:{\cal
H}_\star\to{\cal H}_\star$ given by
    \be
    R:=\rho_0A_0{\cal X},
    \label{R=}
    \end{equation}
and let for each observable $O_\star$ of $Q_\star$,
    \be
    O_\star':=R^\dagger O_\star R.
    \label{R2}
    \end{equation}
Now, inserting (\ref{U-prime-2}) in (\ref{eq90}) and using
(\ref{eq500}), (\ref{eq81b}), (\ref{R=}), and (\ref{R2}), we have
    \bea
    o&=&U_1^{-1}(A_0{\cal X})^{-1}\rho_0^{-1}O_\star\rho_0
    A_0{\cal X}U_1\nn\\
    &=&U_1^{-1}\tilde\eta_0^{-1}(A_0{\cal
    X})^\dagger\eta_{0+}\rho_0^{-1}O_\star\rho_0
    A_0{\cal X}U_1\nn\\
    &=&U_1^{-1}\tilde\eta_0^{-1}(A_0{\cal
    X})^\dagger\rho_0O_\star\rho_0
    A_0{\cal X}U_1\nn\\
    &=&U_1^{-1}\tilde\eta_0^{-1} R^\dagger\,O_\star R U_1\nn\\
    &=& U_1^{-1}\tilde\eta_0^{-1} O_\star' U_1.
    \label{eq502}
    \eea
Because $R$ is an invertible operator acting in ${\cal H}_\star$,
we can express any Hermitian operator acting in ${\cal H}_\star$
in the form $R^\dagger\,O_\star R$ for some Hermitian operator
$O_\star$. This means that in order to obtain the general form of
the observables $o$ we can adopt the general form (\ref{O-star})
for the operator $O'_\star$ in (\ref{eq502}).

Incidentally, recall that we can reproduce the results obtained
for the stationary Klein-Gordon-type fields by taking ${\cal
V}=U(t,t_0)^{-1}$ and ${\cal W}=U(t,t_0)$. This in particular
means that we can recover the results of Section~4.4 by setting
${\cal X}=1$ in the expressions (\ref{eq502}) for $o$. Doing so,
we find that for a stationary field
    \be
    o_{_{\mbox{\footnotesize stat.}}}=
    U_1^{-1}\tilde\eta_0^{-1}\,A_0^\dagger\,\rho_0\;O_\star\;
    \rho_0\,A_0\,U_1.
    \label{eq503}
    \end{equation}
We can also express the observables (\ref{eq502}) for the
nonstationary case in this form provided that we introduce
    \be
    O''_\star:=(\rho_0\,A_0)^{-1\dagger}\,O'_\star\,
    (\rho_0\;A_0)^{-1}.
    \label{eq505}
    \end{equation}
Substituting this equation in (\ref{eq502}) and recalling that
$\rho_0^\dagger=\rho_0$ we find
    \be
    o=U_1^{-1}\,\tilde\eta_0^{-1}\,A_0^\dagger\,\rho_0\;O_\star''\;
    U_1\rho_0\,A_0\, U_1.
    \label{eq506}
    \end{equation}
Again because $\rho_0,A_0:{\cal H}_\star\to{\cal H}_\star$ are
invertible operators, every Hermitian operator acting in ${\cal
H}_\star$ may be expressed in the form (\ref{eq505}). This
together with (\ref{eq503}) and (\ref{eq506}) imply that the
general form of the observables $o$ for a nonstationary
Klein-Gordon-type field is given by the corresponding expressions,
namely (\ref{o-psi-ini-1}) and (\ref{o-psi-ini-2}), for the
stationary case provided that we let $O''_\star$ have the general
form~(\ref{O-star}) and identify the operators ${\cal A}_\pm$
appearing in (\ref{J-K=}) with their value at $t_0$. This yields
the initial values of the field $o\psi$ for a given $\psi\in{\cal
H}$. Obviously, one can no longer obtain an analog of
(\ref{o-psi}) for a nonstationary field. However, the matrix
elements $\cbr\psi,o\phi\ckt_{t_0}$ are still given by the
right-hand side of (\ref{matrix-element}) provided that we set
$D=D_0$ in this equation and identify the operators ${\cal A}_\pm$
with their value at $t_0$ in (\ref{cur-J-K=}).

Finally we wish to point out that because the choice of the
operators ${\cal V},{\cal W}\in G_{\tilde\eta_0}$ is arbitrary,
one can always choose ${\cal V}={\cal W}^{-1}$. In this case,
${\cal X}=1$ and the unitary operator $U'$ is independent of $t$.
This choice simplifies the construction of the Hamiltonians
$\check h$ for the quantum systems $q_{L_{0\pm}}$ from the
Hamiltonians $\check H_\star$ for $Q_\star$. Clearly, for a
$t$-independent $U'$, the Hamiltonians $\check h$ and $\check
H_\star$ are related according to the same formula that relates
the observables $o$ and $O''_\star$.

\section{Quantum Mechanics on the Space of Classical Harmonic
Oscillators}

Consider the equation of motion for a classical harmonic
oscillator with a possibly nonstationary frequency
$\omega=\omega(t)\in\R^+$. This is clearly an example of a
Klein-Gordon-type equation~(\ref{kgt}). In particular if we
consider an isotropic harmonic oscillator in two real (one
complex) dimensions, then we can identify its equation of motion
with the Klein-Gordon-type equation~(\ref{kgt}) corresponding to
the choice:
    \be
    \tilde{\cal H}=\C~\mbox{equipped with the Euclidean
    inner product},~~~~~~~~~D=\omega^2.
    \label{s6-1}
    \end{equation}
The space ${\cal H}$ of solutions of such an oscillator is, as a
vector space, isomorphic to $\C^2$. In fact, because in this case
the Hilbert space ${\cal H}_\star$ is identical with the Euclidean
space $\C^2$, ${\cal H}$ is also isomorphic to the Euclidean space
$\C^2$ as a Hilbert space. Therefore, the classical
oscillators~(\ref{s6-1}) are simple nontrivial systems to which we
can apply the results of Sections~4 and~5 and study their
implications.

Besides being a useful toy model, the classical harmonic
oscillators (\ref{s6-1}) have some application in quantum
cosmology. For instance, the choice
    \be
    \omega=\sqrt\Lambda\,e^{3t}
    \label{frw}
    \end{equation}
corresponds to the Wheeler-DeWitt equation for a flat FRW model
with a positive cosmological constant $\Lambda$, where $t$ is to
be identified with the logarithm of the scale factor,
\cite{abe,simone}.\footnote{This is also true with a slightly more
general expression for $\omega$ for open FRW models.} Therefore,
studying the quantum mechanics of time-dependent oscillators
(\ref{s6-1}) sheds light on some of the basic problems of quantum
cosmology. Specifically, we can use the results of Section~5 to
construct the Hilbert space and the observables for this model,
i.e., solve the Hilbert space problem \cite{isham}.

An interesting outcome of our analysis, which generalizes to other
cosmological models, is that one does not need to obtain solutions
of the Wheeler-DeWitt equation to assess the kinematical structure
of the corresponding quantum cosmological model. The states and
the observables may be formulated in terms of the initial data of
the defining Wheeler-DeWitt equation. The dynamics too may be
expressed in terms of the initial data. However one must first
choose an appropriate Hamiltonian operator. In view of the analogy
with nonrelativistic quantum mechanics, this seems to require
Dirac's canonical quantization of the corresponding classical
system (if there is any) which would assign physical meaning to
the quantum observables and determine (up to factor-ordering
ambiguities) the form of the Hamiltonian. Such a scheme should
naturally be supported by an appropriate correspondence principle.
As we have already imposed the quantum constraint (the
Wheeler-DeWitt equation), the above-mentioned quantization scheme
should be performed on the classical system obtained by imposing
the classical constraint. The flat FRW model with a cosmological
constant is too restricted to allow for a nontrivial classical
internal dynamics. Therefore, we differ a discussion of the above
quantization scheme to the next section where we consider FRW
models coupled to a real scalar field.

In the following, we study the quantum mechanics of general classical
oscillators (\ref{s6-1}) with an arbitrary possibly $t$-dependent
frequency $\omega$.

For the oscillators (\ref{s6-1}), the inner product
(\ref{inv-inn-prod-zero}) takes the form
        \be
        \cbr\psi_1,\psi_2\ckt_{t_0}=\frac{1}{2}\left\{
        L_{0+}[\psi_1^*(t_0)\psi_2(t_0)+\omega_0^{-2}
        \dot\psi_1^*(t_0)\dot\psi_2(t_0)]+
        i\omega_0^{-1}L_{0-}[
        \psi_1^*(t_0)\dot\psi_2(t_0)-
        \dot\psi_1^*(t_0)\psi_2(t_0)]\right\},
        \label{s6-2}
        \end{equation}
    where $t_0$ is an initial value of $t$, $L_{0\pm}$ are a pair of
    real numbers such that $L_{0+}\pm L_{0-}$ are positive, and
    $\omega_0:=\omega(t_0)$. Note that if we choose $L_{0-}=0$ and
    set $L_{0+}=m\omega_0^2$ where $m$ is the mass of the
    oscillator, (\ref{s6-2}) yields the total energy of the
    oscillator at $t_0$. This shows that for a stationary oscillator
    the inner products~(\ref{s6-2}) include the energy inner
    product \cite{energy-inner-product} as a special case.

    The observables $o$ of $q_{L_{0\pm}}$ may be constructed
    as linear combinations (with real coefficients) of a set of four
    basic observables $o_\mu$. These are related via the unitary
    operator $U'$ to the basic observables $\sigma_\mu$ of
    $Q_\star$, where $\mu\in\{0,1,2,3\}$, $\sigma_0$ is the
    $2\times 2$ identity matrix and $\sigma_\mu$ with
    $\alpha\neq 0$ are the Pauli matrices (\ref{pauli}). Letting
    $O_\star''$ of (\ref{eq505}) be equal to $\sigma_\mu$,
    computing the corresponding values for $J_\pm$ and $K_\pm$ of
    (\ref{J-K=}), and using (\ref{o-psi-ini-2}), we
    find, for all $\psi\in{\cal H}$, the following expressions for
    the initial values of $\phi_\mu:=o_\mu\psi$.
        \bea
        \phi_0(t_0)&=&\psi_0,~~~~~~~~~~~~~~~~~~~~~~~~~~~~~~~
        \dot\phi_0(t_0)=\dot\psi_0;
        \label{o0}\\
        \phi_1(t_0)&=&\frac{1}{2}(
        s_+\psi_0+
        is_-\omega_0^{-1}\dot\psi_0),~~~~~~~
        \dot\phi_1(t_0)=\frac{1}{2}(
        i\omega_0s_-\psi_0-s_+\dot\psi_0),
        \label{o1}\\
        \phi_2(t_0)&=&\frac{1}{2}(
        i s_-\psi_0-s_+\omega_0^{-1}\dot\psi_0)
        ,~~~~~~~
        \dot\phi_2(t_0)=\frac{1}{2}(
        -s_+\omega_0\psi_0-is_-\dot\psi_0),
        \label{o2}\\
        \phi_3(t_0)&=&i\omega_0^{-1}\dot\psi_0
        ,~~~~~~~~~~~~~~~~~~~~~~~~\:
        \dot\phi_3(t_0)=-\omega_0\psi_0,
        \label{o3}
        \eea
where
    \begin{equation}
    s_\pm:=
    \frac{{\cal A}_{0+}}{{\cal A}_{0-}}\pm
    \frac{{\cal A}_{0-}}{{\cal A}_{0+}}
    =\left(\frac{L_{0+}+L_{0-}}{L_{0+}-L_{0-}}\right)
    e^{i(\theta_+-\theta_-)}\pm
    \left(\frac{L_{0+}-L_{0-}}{L_{0+}+L_{0-}}\right)
    e^{-i(\theta_+-\theta_-)},
    \label{s=aa}
    \end{equation}
${\cal A}_{0\pm}={\cal A}_\pm(t_0)$, and $\theta_\pm\in\R$ are
arbitrary. As we explain in the Appendix, we can redefine the
basic observables $\sigma_\mu$ in such a way that the phase angles
$\theta_\pm$ disappear from (\ref{o0}) -- (\ref{o3}). Therefore,
the observables $o_\mu$ with $s_\pm$ given by (\ref{s=aa}) and
$\theta_\pm=0$ provide a set of basic observables for
$q_{L_{0\pm}}$.

Next, we construct a Hamiltonian operator $\check h$ acting in
${\cal H}$. Following the prescription described in the last
paragraph of Section~5.4, $\check h$ may be obtained from a
Hamiltonian $\check H_\star$ acting in ${\cal H}_\star$ by the
same unitary transformation $U'$ that is used to construct the
basic observables $o_\mu$. This in turn means that $\check h$ will
be a linear combination of $o_\mu$ with real coefficients.

If the oscillator is stationary, i.e., $\omega$ does not depend on
$t$, we have the canonical Hamiltonian $h$ that is obtained via
the unitary transformation $U$ of (\ref{UU}) from the Hamiltonian
$H_\star$ of (\ref{H-star}). The latter has the simple form
    \be
    H_\star=\hbar\omega\sigma_3.
    \label{H-star-oss}
    \end{equation}
As we discussed in section~4.3, the Hamiltonian $h$ identifies the
parameter $t$ with time. In view of (\ref{H-star-oss}), $H_\star$
and consequently $h$ describe the interaction of a spin-half
particle with a constant magnetic field~\cite{nova}.

For a nonstationary oscillator, where $\omega$ depends on $t$, we
can still choose
    \be
    \check H_\star=H_\star.
    \label{rot}
    \end{equation}
But the corresponding Hamiltonian $\check h$ does not generate
$t$-translations in ${\cal H}$. Instead, it defines its own
time-parameter $\check t\in\R$. Clearly, in this case, $\check
h=\hbar\omega o_3$. Therefore, the eigenvectors of $\check h$ do
not depend on $\check t$ and its evolution operator is readily
obtained as \cite{nova}
    \be
    \check\upsilon (\check t,\check t_0)=
    e^{-i\Omega(\check t,\check t_0)\,o_3},
    \label{evo-op}
    \end{equation}
where
    \be
    \Omega(\check t,\check t_0):=\int_{\check t_0}^{\check t}
    \omega(t)dt,
    \label{Omega}
    \end{equation}
and $\check t_0$ is an initial time. Moreover, because $\check h$
is related to $\check H_\star$ via the unitary operator $U'$,
$\check\upsilon(\check t,\check t_0)$ may be obtained from the
time-evolution operator for $\check H_\star$, namely
    \[\check U_\star(\check t,\check t_0)=
    e^{-i\Omega(\check t,\check t_0)\,\sigma_3}=
    \cos[\Omega(\check t,\check t_0)]\,\sigma_0-i
    \sin[\Omega(\check t,\check t_0)]\,\sigma_3.\]
This yields
    \be
    \check\upsilon(\check t,\check t_0)=
    \cos[\Omega(\check t,\check t_0)]\,o_0-i
    \sin[\Omega(\check t,\check t_0)]\,o_3.
    \label{c-U}
    \end{equation}

Now, let $\psi_{\check t}\in{\cal H}$ be an evolving state vector
with initial condition $\phi\in{\cal H}$, i.e.,
    \be
    \psi_{\check t}=\check\upsilon(\check t,\check t_0)\phi.
    \label{psi=u-psi}
    \end{equation}
Then, in view of (\ref{o0}), (\ref{o3}) and (\ref{c-U}), the
initial data $(\psi_{\check t}(t_0), \dot\psi_{\check t}(t_0))$
for $\psi_{\check t}$ are related to the initial data
$(\phi(t_0),\dot\phi(t_0))$ for $\phi$ according to
    \bea
    \psi_{\check t}(t_0)&=&\cos[\Omega(\check t,\check t_0)]\,
    \phi(t_0)+\omega_0^{-1} \sin[\Omega(\check t,\check
    t_0)]\,\dot\phi(t_0),
    \label{ini-1}\\
    \dot\psi_{\check t}(t_0)&=&i\omega_0
    \sin[\Omega(\check t,\check t_0)]\phi(t_0)+
    \cos[\Omega(\check t,\check t_0)]\,\dot\phi(t_0).
    \label{ini-2}
    \eea
Clearly, the time-evolution generated by the Hamiltonian
(\ref{rot}) corresponds to `rotations' of the state vectors in ${\cal
H}$. The time-parameter $\check t$ of this Hamiltonian cannot be
identified with the parameter $t$ of the defining classical
oscillator.

The description of the dynamics generated by the Hamiltonian
(\ref{rot}) in terms of rotations in ${\cal H}$ generalizes to
arbitrary choices for the Hamiltonian $\check h$. This is because
any Hamiltonian $\check H_\star$ belongs to the Lie algebra
$u(2)$. As the dynamics of the states take place in the projective
Hilbert space \cite{nova}, which in this case has the structure of
$\C P^1$, it is the traceless part of $\check H_\star$ that is
physically significant. This belongs to the Lie algebra $su(2)$.
Hence the group $SU(2)$ serves as the dynamical Lie group for the
quantum systems of $Q_\star$ and consequently $q_{L_{o\pm}}$. In
particular, the evolution operator for any Hamiltonian $\check h$
corresponds to a ($SU(2)$-) rotation of the states in the
Schr\"odinger-picture (respectively of observables in the
Heisenberg-picture).

\section{Quantum Mechanics of a FRW Model Coupled to a Real
Scalar Field}

It is well-known that the canonical quantization of an FRW model
coupled to a real scalar field $\varphi$ yields a single quantum
constraint \cite{isham,page}:
    \be
    \hat{\cal K}\psi=0,
    \label{wdw-00}
    \end{equation}
where $\hat{\cal K}$ is obtained by quantizing the classical
Hamiltonian constraint \cite{blyth-isham}:
    \be
    {\cal K}:=-\pi_\alpha^2+\pi_\varphi^2-\kappa\,
    e^{4\alpha}+e^{6\alpha}\,V(\varphi)=0,
    \label{c-constraint}
    \end{equation}
$\pi_\alpha$ and $\pi_\varphi$ are respectively the momenta
associated with the logarithm $\alpha$ of the scale factor $a$ and
the scalar field $\varphi$, $V=V(\varphi)$ is a real-valued
potential for the field $\varphi$, $\kappa=-1,0,1$ determines
whether the universe is open, flat, or closed, respectively, and
we have used the natural system of units as discussed in
\cite{wiltshire}.

The above-mentioned quantization of the classical
constraint~(\ref{c-constraint}) means the canonical quantization
$\Xi$ that respectively associates to the classical unconstrained
phase-space variables $\alpha, \varphi, \pi_\alpha$, and
$\pi_\varphi$ the operators $\Xi(\alpha),\Xi(\varphi),
\Xi(\pi_\alpha)$, and $\Xi(\pi_\varphi)$ that act in the auxiliary
(kinematic) Hilbert space ${\cal H}'=L^2(\R^2)$, where $\R^2$ is
the configuration space parameterized by $(\alpha,\varphi)$.
Specifically, one treats $\vec X:=(\Xi(\alpha),\Xi(\varphi))$ as
the position operator and $\vec
P:=(\Xi(\pi_\alpha),\Xi(\pi_\varphi))$ as the momentum operator
acting in $L^2(\R^2)$ and uses the position kets $|\vec
x\kt=:|\alpha,\varphi\pkt$ to express the quantum
constraint~(\ref{wdw-00}) in terms of the wave functions
    \be
    \psi(\alpha,\varphi):=\pbr\alpha,\varphi|\psi\pkt,
    \label{wf-00}
    \end{equation}
where $\pbr\cdot|\cdot\pkt$ stands for the inner product in ${\cal
H'}=L^2(\R^2)$. In view of the identities
    \bea
    \pbr\alpha,\varphi|\:\Xi(\alpha)&=&\alpha\pbr\alpha,\varphi|,
    ~~~~~~~\pbr\alpha,\varphi|\:\Xi(\varphi)=
    \varphi\pbr\alpha,\varphi|,\nn\\
    \pbr\alpha,\varphi|\:\Xi(\pi_\alpha)&=&-i\frac{\partial}{\partial
    \alpha}\pbr\alpha,\varphi|,~~~~~~~
    \pbr\alpha,\varphi|\:\Xi(\pi_\varphi)=-i\frac{\partial}{\partial
    \varphi}\pbr\alpha,\varphi|,\nn
    \eea
the quantum constraint~(\ref{wdw-00}) --- written in the form
$\pbr\alpha,\varphi|\hat{\cal K}\psi\pkt=0$ --- yields the
Wheeler-DeWitt equation~(\ref{wdw-0}):
    \be
    \left[ -\frac{\partial^2}{\partial\alpha^2}+
    \frac{\partial^2}{\partial\varphi^2}+
    \kappa\,e^{4\alpha}-e^{6\alpha}V(\varphi)\right]\,
    \psi(\alpha,\varphi)=0.
    \label{wdw-0-2}
    \end{equation}
Here and in what follows we set $\hbar=1$, but recover $\hbar$
where appropriate.

As we have noted in Section~1, the Wheeler-DeWitt
equation~(\ref{wdw-0-2}) is another example of Klein-Gordon-type
equations. We can write it in the form~(\ref{kgt}), if we identify
$\alpha$ with the variable $t$ and let $\tilde{\cal H}=L^2(\R)$
and
    \be
    D=-\frac{\partial^2}{\partial\varphi^2}+e^{6\alpha}\, V(\varphi)
    -\kappa\,e^{4\alpha}.
    \label{D=wdw-0}
    \end{equation}
Clearly $D$ is a Hermitian operator acting in $L^2(\R)$. However,
depending on the form of $V$ and the value of $\kappa$ it may or
may not be positive-definite. In what follows we shall only
consider nonnegative confining potentials, i.e., suppose that for
all $\varphi\in\R$, $V(\varphi)\geq 0$ and
$\lim_{\varphi\to\pm\infty}V(\varphi)=+\infty$. In this case $D$
has a nondegenerate discrete spectrum \cite{messiyah}. Moreover,
for the open and flat models, $D$ is a positive-definite operator.

A typical choice for the potential $V$ is
    \be
    V(\varphi)=m^2\varphi^2
    \label{massive}
    \end{equation}
which corresponds to $\varphi$ being a massive scalar field of
mass $m$. For such a field, the operator $D$ is identical with the
Hamiltonian operator of a time-dependent quantum harmonic
oscillator. Therefore its eigenvalue problem can be easily solved
\cite{jmp-98}.

Before we apply the theory developed in the preceding sections to
formulate a quantum cosmology based on the Wheeler-DeWitt
equation~(\ref{wdw-0-2}), we wish to point out that the definition
of the operator $\hat{\cal K}$ alternatively (\ref{wdw-00})
suffers from a factor-ordering ambiguity. For example, one may
quantize the term $\pi_\alpha^2$ appearing in the classical
constraint (\ref{c-constraint}) according to
$\Xi(\pi_\alpha^2)=\Xi(\alpha)^p\;\Xi(\pi_\alpha)\;\Xi(\alpha)^{-2p}
\;\Xi(\pi_\alpha)\;\Xi(\alpha)^p$ for any $p\in\R$. This leads to
additional terms in the left-hand side of the Wheeler-DeWitt
equation~(\ref{wdw-0-2}). The factor-ordering problem is more
transparent if one expresses the classical constraint in terms of
the scale factor $a$. Certainly it does not go away if one uses
$\alpha$.

The presence of $\alpha$ in the expression for $D$ indicates that
the corresponding Wheeler-DeWitt equation is a nonstationary
Klein-Gordon-type equation. We can apply the results of Sections~3
and~5 to obtain the most general Hilbert space structure on the
physical space ${\cal H}$ of solutions of the Wheeler-DeWitt
equation, construct the observables, and perhaps more importantly
reduce the problem of the identification of time to the issue of
choosing a Hamiltonian operator acting in ${\cal H}$ (equivalently
acting in ${\cal H}_\star$). Following the approach of Section~6,
as a method of choosing a specific Hamiltonian, we propose to
perform a canonical quantization of the classical system obtained
after imposing the classical constraint.

According to the results of Section~3, all the choices for an
inner product on the physical Hilbert space ${\cal H}$ are
physically equivalent. We can choose the inner product
(\ref{inv-inn-prod-zero}) with $D_0=D(\alpha_0)$ for any
$\alpha_0\in\R$ for the flat and open FRW models. For a closed FRW
model we may identify $D_0$ with $D(\alpha_0)$ for any $\alpha_0$
that makes $D(\alpha_0)$ positive-definite. For a positive
confining potential $V$ the smallest eigenvalue of $D+\kappa
e^{4\alpha}$ is necessarily positive. This together with the
particular $\alpha$-dependence of $D$ imply that there is always
some $\alpha_0\in\R$ such that $D(\alpha_0)$ is a
positive-definite operator. In the following we shall choose such
an $\alpha_0$ and endow ${\cal H}$ with an inner product of the
form (\ref{inv-inn-prod-zero}).

In order to facilitate the application of the results of Section~5
and avoid potential ambiguities, we introduce for each real number
$\alpha$ and Wheeler-DeWitt field $\psi$ the value $\psi(\alpha)$
of $\psi$. This is the function:
    \[ \psi(\alpha)[\varphi]:=\psi(\alpha,\varphi),~~~~~~~~~
    \forall\varphi\in\R.\]
We shall view $\psi(\alpha)$ as an element of the Hilbert space
$\tilde{\cal H}=L^2(\R)$.\footnote{This is a standard approach in
dealing with hyperbolic partial differential equations. See for
example \cite{evans}.} In this way we can rewrite the
Wheeler-DeWitt equation~(\ref{wdw-0-2}) in the standard
form~(\ref{kgt}) of a Klein-Gordon-type equation, namely
    \be
    \ddot\psi(\alpha)+D\psi(\alpha)=0,
    \label{wdw}
    \end{equation}
where a dot means an $\alpha$-derivative and the operator $D$ is
viewed as a linear operator acting in the abstract Hilbert space
$\tilde{\cal H}=L^2(\R)$.

Because the inner product (\ref{inv-inn-prod-zero}) is
positive-definite, unlike the Klein-Gordon inner product
\cite{isham,page}, it allows for a genuine probability
interpretation for the Wheeler-DeWitt fields. Evidently such an
interpretation is useful only if we also have a set of basic
observables. As we discussed in detail in Section~5, the
observables of the quantum mechanics $q_{L_{0\pm}}$ having ${\cal
H}$ with inner product (\ref{inv-inn-prod-zero}) as its Hilbert
space are obtained via the unitary transformations $U'$ from the
observables $O_\star''$ of $Q_\star$.

It is not difficult to see that any observable $O_\star''$ can be
expressed in terms of the following basic observables:
    \be
    O_\mu:=1\otimes\sigma_\mu,~~~~~~~~~~~~~~~~~~~~
    \hat Q:=\hat{\tilde q} \otimes\sigma_0,
    ~~~~~~~~~~~~~~~~~~~~
    \hat P:=\hat{\tilde p} \otimes\sigma_0,
    \label{basic-o}
    \end{equation}
where $1$ stands for the identity operator of $\tilde{\cal H}$,
$\mu\in\{0,1,2,3\}$, $\sigma_0$ is the $2\times 2$ identity
matrix, $\sigma_\mu$ with $\alpha\neq 0$ are the Pauli matrices
(\ref{pauli}), and $\hat{\tilde q}$ and $\hat{\tilde p}$
respectively play the role of the position and momentum operators
acting in $\tilde{\cal H}=L^2(\R)$. In particular, they satisfy
the canonical commutation relations $[\hat{\tilde q},\hat{\tilde
p}]=i\hbar 1$.

We can represent the elements of $\tilde{\cal H}$ in the position
basis. The position kets $|q\kt$, with $q\in\R$, are defined by
$\hat{\tilde q}|q\kt=q|q\kt$. They satisfy
    \[\br q|q'\kt=\delta(q-q'),~~~~~~~~
    \br q|\hat{\tilde p}=-i\hbar\,\frac{d}{dq}\;\br q|.\]

As seen from (\ref{basic-o}), any linear combination of $O_\mu$
commutes with both $\hat Q$ and $\hat P$. In particular
$\{O_3,\hat Q\}$ is a smallest set of commuting basic observables.
Therefore, we may use their generalized eigenvectors\footnote{Here
`generalized eigenvector' means a generalized eigenfunction
(distribution) viewed as an abstract vector. It does not refer to
the concept of a generalized eigenvector used in linear algebra.}
to construct a complete basis of ${\cal H}_\star$. They are given
by $\xi^{(q,\epsilon)}:=|q\kt\otimes e_\epsilon$ where
$\epsilon=\pm $ and
    \[e_+:=\left(\begin{array}{c}
        1\\
        0\end{array}\right),~~~~~~~~~~~~~~~~
      e_-:=\left(\begin{array}{c}
        0\\
        1\end{array}\right).\]
Clearly, the following orthonormality and completeness relations
hold:
    \be
    \br\xi^{(q,\epsilon)},\xi^{(q',\epsilon')}\kt=\delta_{\epsilon,\epsilon'}
    \delta(q-q'),~~~~~~~~~~~~~~~~~
    \sum_{\epsilon=\pm}\int_{-\infty}^{\infty} dq\:
    |\xi^{(q,\epsilon)}\kt\br\xi^{(q,\epsilon)}|=O_0.
    \label{ortho-comp}
    \end{equation}
As a result, any two-component state vector $\xi\in\tilde{\cal
H}\otimes\C^2$ may be represented by the {\em wave functions}
$g(q,\epsilon):= \br\xi^{(q,\epsilon)},\xi\kt$ according to
    \[ \xi=\sum_{\epsilon=\pm}\int_{-\infty}^{\infty}dq\:
        g(q,\epsilon)\;\xi^{(q,\epsilon)}.\]

In the following we will, without loss of generality,  adopt
 ${\cal X}=1$ in the expression (\ref{U-prime-2}) for
the operator $U'$. This choice identifies $U'$ with the value of
the operator $U$ of (\ref{UU}) at $t=t_0=\alpha_0$ and, as we
explained in Section~5.4, simplifies the construction of the
Hamiltonian operators acting in ${\cal H}$.

We can use the operator $U'$ to define a set of basic observables
for $q_{L_{0\pm}}$. These are given by substituting the
observables (\ref{basic-o}) for $O_\star''$ in (\ref{eq506}). We
will denote the observables of  $q_{L_{0\pm}}$ obtained in this
way from $O_\mu,\hat Q$ and $\hat P$ by $o_\mu,\hat q$ and $\hat
p$, respectively. In order to determine the latter we express the
former in the matrix form (\ref{O-star}), i.e., find the
corresponding operators $\tilde O_1, \tilde O_2$ and $\tilde{\cal
O}$, and use (\ref{o-psi-ini-1}) -- (\ref{J-K=}) with $D$ and
${\cal A}_\pm$ replaced with their value at $\alpha_0$.

The action of $o_\mu,\hat q$ and $\hat p$ on an arbitrary solution
$\psi$ of the Wheeler-DeWitt equation is described as follows.
$\phi_\mu:=o_\mu\psi$ is determined by the initial conditions
given by (\ref{o0}) -- (\ref{o3}). $\phi_{\hat q}:= \hat q\,\psi$
is determined by the initial conditions
    \be
    \phi_{\hat q}(\alpha_0)=\hat{\tilde q}\:\psi_0,
    ~~~~~~~~~~~~~~~~~~
    \dot\phi_{\hat q}(\alpha_0)=\hat{\tilde q}\:\dot\psi_0.
    \label{act-phi}
    \end{equation}
Similarly, $\phi_{\hat p}:=\hat p\,\psi$ is determined by the
initial conditions:
    \be
    \phi_{\hat p}(\alpha_0)=\hat{\tilde p}\:\psi_0,
    ~~~~~~~~~~~~~~~~~~
    \dot\phi_{\hat p}(\alpha_0)=\hat{\tilde p}\:\dot\psi_0.
    \label{act-pi}
    \end{equation}

Obviously, the eigenvectors
    \be
    \psi^{(q,\epsilon)}:= U^{'-1}\xi^{(q,\epsilon)}
    \label{161.5}
    \end{equation}
of the commuting observables $o_3$ and $\hat q$ form a complete
orthonormal basis for ${\cal H}$:
    \be
    \cbr\psi^{(q,\epsilon)},
    \psi^{(q',\epsilon')}\ckt_{\alpha_0}=
    \delta_{\epsilon,\epsilon'}
    \delta(q-q'),~~~~~~~~~~~~~~~~~
    \sum_{\epsilon=\pm}\int_{-\infty}^{\infty} dq\:
    |\psi^{(q,\epsilon)}\ckt\cbr\psi^{(q,\epsilon)}|=I,
    \label{ortho-comp-psi}
    \end{equation}
where for all $\psi,\phi\in{\cal H}$ the operator
$|\psi\ckt\cbr\phi|$ is defined by
    \[|\psi\ckt\cbr\phi|\varrho:=
    \cbr\phi,\varrho\ckt_{\alpha_0}\psi,\]
$\varrho\in{\cal H}$ is an arbitrary state vector, and $I$ denotes
the identity operator for ${\cal H}$. Clearly, the generalized
eigenvectors $\psi^{(\varphi,\epsilon)}$ represent a complete set
of {\em localized states} of the corresponding FRW universe.

Furthermore, any solution $\psi\in{\cal H}$ of the Wheeler-DeWitt
equation~(\ref{wdw-0-2}) may be expressed as
    \be
    \psi=\sum_{\epsilon=\pm}\int_{-\infty}^{\infty}dq\:
        f(q,\epsilon)\:\psi^{(q,\epsilon)}.
    \label{wf-0}
    \end{equation}
where
    \be
    f(q,\epsilon):=
    \cbr\psi^{(q,\epsilon)},\psi\ckt_{\alpha_0}=
    \br\xi^{(q,\epsilon)},U'\psi\kt
    =\frac{1}{2}\left[
    \br q|{\cal A}_{0\epsilon}\psi_0\kt+
    \epsilon i \br q|{\cal A}_{0\epsilon}
    D_0^{-1/2}\dot\psi_0\kt\right].
    \label{wf}
    \end{equation}
In the derivation of this expression we have made use of
(\ref{U-psi}) which yields the operator $U'$ upon evaluating its
right-hand side at $\alpha_0$. We also recall that ${\cal
A}_{0\pm}$ appearing in (\ref{wf}) are arbitrary invertible
operators commuting with $D_0$. They reflect the freedom of the
choice of the operators $L_{0\pm}$ in the expression for the inner
product~(\ref{inv-inn-prod-zero}) on ${\cal H}$. For instance, we
can set, without loss of generality, ${\cal A}_{0+}={\cal
A}_{0+}=(\ell\, D_0)^{1/2}$ where $\ell$ is a positive real
constant. Then in view of (\ref{L=2}), $L_{0+}=\ell\,D_0$,
$L_{0-}=0$, and (\ref{inv-inn-prod-zero}) turns into an energy
inner product \cite{energy-inner-product}.

Having obtained position- and momentum-like operators acting in
${\cal H}$, we can also define a set of coherent states. The
latter will be represented by the eigenvectors of the annihilation
operator $\hat {\rm a}:=\sqrt{k/2\hbar}(\hat q+i\hat p/k)$,
where $k\in\R^+$ is a constant.\footnote{We may also
consider `directed coherent states' where the corresponding state
vectors are the eigenvectors of the `directed annihilation
operators': $\hat {\rm a}_{\vec n}:=\sqrt{k/2\hbar}(\hat q+i\hat
p/k)\sum_{i=1}^3 n_i o_i$, where $\vec n=(n_1,n_2,n_3)\in\R^3$ is
a unit vector.}

By construction, the wave function (\ref{wf}), which maps
$\R\times\{-1,1\}$ into $\C$, determines the solution $\psi$ of
the Wheeler-DeWitt equation uniquely. In terms of the wave
functions~(\ref{wf}), the inner product of a pair of
Wheeler-DeWitt fields $\psi_1,\psi_2\in{\cal H}$ takes the form
    \be
    \cbr\psi_1,\psi_2\ckt_{\alpha_0}=\br U'\psi_1,U'\psi_2\kt=
    \sum_{\epsilon=\pm}\int_{-\infty}^{\infty}dq\:
        f_1(q,\epsilon)^* f_2(q,\epsilon),
    \label{inner-wf}
    \end{equation}
where $f_i$ denotes the wave function (\ref{wf}) associated with
$\psi_i$.

The observables $o$ of $q_{L_{0\pm}}$ are also uniquely specified
in terms of their representation in the basis
$\{\psi^{(q,\epsilon)}\}$; for all $\psi\in{\cal H}$
    \be
    o\,\psi=\sum_{\epsilon=\pm}\int_{-\infty}^{\infty}dq\:
    [\hat o\,f(q,\epsilon)]\;\psi^{(q,\epsilon)},
    \label{observables-wf}
    \end{equation}
where
    \bea
    \hat o\:f(q,\epsilon)&=&
    \sum_{\epsilon'=\pm}\int_{-\infty}^{\infty}dq'\:
    o(q,\epsilon;q',\epsilon')\:f(q',\epsilon'),
    \label{wf-10}\\
    o(q,\epsilon;q',\epsilon')&:=&
    \cbr\psi^{(q,\epsilon}),o\,\psi^{(q',\epsilon')}
    \ckt_{\alpha_0}.
    \label{wf-11}
    \eea
If we let $O:=U^{'}o\:U^{'-1}$, then we can express the matrix
elements $o(q,\epsilon;q',\epsilon')$ of the operator $o$ in the
form
    \be
    o(q,\epsilon;q',\epsilon')=
    \br\xi^{(q,\epsilon)},O\xi^{(q',\epsilon')}
    \kt=\br q|O_{\epsilon,\epsilon'}|q'\kt,
    \label{kernel}
    \end{equation}
where $O_{\epsilon,\epsilon'}:=e_\epsilon^\dagger O\,
e_{\epsilon'}$. Similarly we have, for any pair of
Wheeler-DeWitt fields $\psi_1$ and $\psi_2$, with wave
functions $f_1$ and $f_2$, and observables
$o:{\cal H}\to{\cal H}$,
    \be
    \cbr\psi_1,o\psi_2\ckt_{\alpha_0}=
     \sum_{\epsilon=\pm}\int_{-\infty}^{\infty}dq\:
        f_1(q,\epsilon)^* \hat o\,f_2(q,\epsilon),
    \label{exp-va-wf}
    \end{equation}
In particular, we can compute the expectation values of the
observable $o$ using the wave functions $f$.

The above discussion shows that we can formulate the quantum
cosmology of any FRW model coupled to a real scalar field in terms
of the wave functions~(\ref{wf}). The latter belong to the Hilbert
space $L^2(\R\times\{-1,1\})$ which is isomorphic to
$L^2(\R)\oplus L^2(\R)={\cal H}_\star$. The observables $o$
(including any Hamiltonian operator) are specified by the
kernels~(\ref{kernel}) according to (\ref{observables-wf}) and
(\ref{wf-10}) or alternatively by the operators $\hat o$ that act
on the wave functions. For instance, for the basic observables
$o_3$, $\hat q$ and $\hat p$, we have
    \be
    \hat o_3\,f(q,\epsilon)=\epsilon\,f(q,\epsilon),
    ~~~~~~~
    \hat{\hat q}\;f(q,\epsilon)=q\;f(q,\epsilon),~~~~~~~
    \hat{\hat p}\;f(q,\epsilon)=-i\hbar
    \frac{\partial}{\partial q}\,f(q,\epsilon).
    \label{wf-20}
    \end{equation}

In view of the above description of the wave functions
$f(q,\epsilon)$, the name `{\em wave function of the universe}'
seems more appropriate for $f(q,\epsilon)$ than the wave functions
$\psi(\alpha,\varphi)$ of (\ref{wf-00}), provided that we assign a
physical meaning for the variables $q$ and $\epsilon$.

It is worthy of noting that the physical meaning of the variables
$\alpha$ and $\varphi$ of the wave functions
$\psi(\alpha,\varphi)$ is not clear. By definition, these
variables are respectively the eigenvalues of the position-like
operators $\Xi(\alpha)$ and $\Xi(\varphi)$. Because these
operators act in the non-physical auxiliary (kinematic) Hilbert
space and do not commute with the operator $\hat{\cal K}$, they
cannot be restricted to the physical Hilbert space ${\cal H}$.
This in turn means that they do not represent physical
observables, and their eigenvalues have nothing to do with the
results of any physical measurement. Therefore, {\em a priori} one
does not have a good reason for identifying the independent
variables $\alpha$ and $\varphi$ of the wave functions
$\psi(\alpha,\varphi)$ with the classical counterparts of
$\Xi(\alpha)$ and $\Xi(\varphi)$ --- which are nevertheless
denoted by the same symbols. To the author's best knowledge the
only way of relating these variables to the classical observables
is by restricting to the approximate WKB solutions of the Wheeler-DeWitt
equation \cite{bryce-1,vilenkin-89,wiltshire}. The main purpose of
the present study is to formulate a quantum theory of cosmology
that avoids this and similar restrictions. Hence we hold that at a
fundamental level, the variables $\alpha$ and $\varphi$ appearing
in $\psi(\alpha,\varphi)$ and therefore the wave functions
$\psi(\alpha,\varphi)$ lack a clear physical interpretation. As we
shall see below, the same is not the case for the variables $q$
and $\epsilon$ and the wave functions $f(q,\epsilon)$. It turns
out that the physical meaning of the latter is intertwined with
the dynamical aspects of the theory.

Consider a linear operator $\hat h'$ acting in the space
$L^2(\R)\oplus L^2(\R)$ of the wave functions~(\ref{wf}) and
corresponding to a Hamiltonian operator $h'$ acting on the
Wheeler-DeWitt fields $\psi$ of (\ref{wdw-0-2}). It is not
difficult to check that because $h':{\cal H}\to{\cal H}$ is a
Hermitian operator, so is $\hat h':L^2(\R)\oplus L^2(\R)\to
L^2(\R)\oplus L^2(\R)$. Moreover, every evolving Wheeler-DeWitt
field $\psi_{\check t}$ satisfying the Schr\"odinger equation
    \be
    i\hbar\frac{d}{d\check t}\,
    \psi_{\check t}=h'\,\psi_{\check t},
    \label{sch-eq-wf-wdw}
    \end{equation}
may be specified in terms of its wave function $f_{\check t}
(q,\epsilon)=:f(q,\epsilon;{\check t})$ that fulfills the
Schr\"odinger equation
    \be
    i\hbar\frac{\partial}{\partial\check t}\,
    f(q,\epsilon;\check t)=\hat h'\,f(q,\epsilon;\check t).
    \label{sch-eq-wf}
    \end{equation}
This reduces the study of the dynamical aspects of the quantum FRW
models with a real scalar field to the formulation of a canonical
quantization of an appropriate classical system that would assign
physical meaning to the operators $\hat{\hat q}$ and $\hat o_3$
(and consequently to the variables $q$ and $\epsilon$ and the operators
${\hat q}$ and $o_3$) and leads to a choice of a Hermitian
operator $\hat h'$ acting in the Hilbert space $L^2(\R)\oplus
L^2(\R)$.

Before presenting further details of the dynamical structure of
the theory, we wish to comment on the factor-ordering ambiguity
associated with the definition of the Wheeler-DeWitt equation. As
we explained above, we can formulate the whole theory in terms of
the wave functions~(\ref{wf}). These are the coefficients of the
Wheeler-DeWitt fields in the position-basis (\ref{161.5}). If one
chooses another factor-ordering prescription for the operator
corresponding to the Hamiltonian constraint, one obtains a
Wheeler-DeWitt equation which differs from (\ref{wdw-0-2}).
However, the latter will be a linear hyperbolic partial
differential equation with the same highest order terms; the
corresponding differential operators have identical leading symbol
\cite{nakahara}. This in turn implies that although a different
factor-ordering leads to a different form of the solutions for a
given initial data, the space of solutions will have the same
vector space structure. In particular, the position-basis vectors
(\ref{161.5}) will be different, but one will have the same set of
wave functions (\ref{wf}). As the Hilbert space ${\cal H}$ is
uniquely determined (up to unitary-equivalence) by the vector
space structure of the space of solutions, the factor-ordering
problem is not relevant to the kinematical structure of the
quantum theory. This argument relies on two basic assumptions:
    \begin{enumerate}
    \item It is the space of solutions of the Wheeler-DeWitt
    equation that serves as the Hilbert space of the quantum
    theory.
    \item The Hilbert space is separable and therefore its Hilbert
    space structure is unique \cite{reed-simon}.
    \end{enumerate}
The irrelevance of the factor-ordering problem for the kinematics
of the quantum theory holds generally for any quantum
gravitational model that does not violate these assumptions.

Next, we wish to explore the dynamical aspects of the quantum FRW
model defined by the Wheeler-DeWitt equation~(\ref{wdw-0-2}) that
has the classical theory defined by the classical Hamiltonian
constraint~(\ref{c-constraint}) as its `classical limit.' The
first step in this direction is to identify the physical meaning
of the configuration variables $q$ and $\epsilon$ as well as the
time variable $\check t$ appearing as the argument of an evolving
wave function $f(q,\epsilon;\check t)$.

Using the prescription provided by Dirac's canonical quantization,
$q$ and $\epsilon$ are to be identified with the coordinates of
the classical configuration space. The fact that the wave
functions determine the solutions of the Wheeler-DeWitt equation
(i.e., the quantum constraint) suggests that $q$ must be chosen
among the configuration variables of the classical system $C_0$
obtained after imposing the classical constraint. The same applies
to the $\epsilon$ degree of freedom whose interpretation is
slightly more subtle.

In light of analogy with nonrelativistic quantum mechanics, we
propose to identify the time parameter $\check t$ with a classical
time parameter for $C_0$. One has the well-known set of choices
corresponding to the internal and external classical time
variables \cite{isham}. For example if we identify $\check t$ with
the variable $\alpha$ appearing in the classical Hamiltonian
constraint (\ref{c-constraint}), we obtain the corresponding
classical Hamiltonian as outlined in Ref.~\cite{blyth-isham}: We
first solve (\ref{c-constraint}) for $\pi_\varphi$ and substitute
the result in the expression for the first-order Lagrangian
    \be
    {\cal L}:=\pi_\alpha\dot\alpha+ \pi_\varphi\dot\varphi+
    N {\cal K},
    \label{lag-1}
    \end{equation}
where $N$ is the lapse function. Employing $\alpha=\check t$ and
imposing the classical constraint ${\cal K}=0$, we then find
    \be
    {\cal L}=\pi_\varphi\dot\varphi\pm
    \sqrt{\pi_\varphi^2-\kappa\,
    e^{4\alpha}+e^{6\alpha}\,V(\varphi)}.
    \label{lagrangian}
    \end{equation}
We may also describe the classical dynamics of the resulting
system using a Hamiltonian formulation based on the classical
Hamiltonian
    \be
    {\cal K}_0:=\pm\sqrt{\pi_\varphi^2-\kappa\,
    e^{4\alpha}+e^{6\alpha}\,V(\varphi)}.
    \label{K-zero}
    \end{equation}

The expression~(\ref{lagrangian}) suggests the natural
identification of $q$ with $\varphi$. The undetermined sign in
(\ref{lagrangian}) and (\ref{K-zero}) is usually fixed by
demanding that the Hamiltonian (\ref{K-zero}) be positive
\cite{blyth-isham}. This sign may be viewed as a suitable
candidate for the classical analog of the sign $\epsilon$
appearing as one of the arguments of the wave functions
(\ref{wf}). As it multiplies the Hamiltonian, it may be connected
to whether one takes $\alpha$ or $-\alpha$ as a time-variable.
Both these choices are necessary whenever the dynamics of the
classical universe involves a recollapse. The quantum description
of the model seems to include this phenomenon whether the
classical universe recollapses or not.

Incorporating the $\epsilon$ degree of freedom as the sign of the
rate of change of $\alpha$ (with respect to any classical physical
time $\tau$) in the above choice of an internal time is equivalent to
setting
    \be
    \check t=\epsilon\alpha,
    \label{class-time}
    \end{equation}
where $\epsilon:={\rm sign}(d\alpha(\tau)/d\tau)$, sign$(0):=1$,
and sign$(x):=x/|x|$ for $x\neq 0$. Repeating the above derivation
of the Lagrangian (\ref{lagrangian}) and the Hamiltonian
(\ref{K-zero}) with the choice~(\ref{class-time}) for time and
requiring that the resulting Hamiltonian be positive yield
    \bea
    {\cal L}&=&\pi_\varphi\dot\varphi+
    \sqrt{\pi_\varphi^2-\kappa\,
    e^{4\epsilon\check t}+e^{6\epsilon\check t}\,V(\varphi)},
    \label{lagrangian-2}\\
    {\cal K}_0&:=&\sqrt{\pi_\varphi^2-\kappa\,
    e^{4\epsilon\check t}+e^{6\epsilon\check t}\,V(\varphi)}.
    \label{K-zero-2}
    \eea

The above discussion shows that the structure of the classical
system $C_0$ provides the physical meaning of the variables
$q,\epsilon$ and $\check t$. The next step is to perform a
canonical quantization of the system $C_0$ to clarify the physical
interpretation of the operators $\hat{\hat q}$ and $\hat o_3$
(respectively $\hat q$ and $o_3$) and construct the Hamiltonian
$\hat h'$ (respectively $h'$).

It is important to note the difference between the quantum
operators $\Xi(\varphi)\equiv\varphi$ and $\Xi(\pi_\varphi)\equiv
-i\partial/\partial\varphi$ that appear in the Wheeler-DeWitt
equation~(\ref{wdw-0-2}) and $\hat{\hat q}$ and $\hat{\hat p}$ of
(\ref{wf-20}). The former are obtained by quantizing $\varphi$ and
$\pi_\varphi$ viewed as classical observables of the system before
imposing the classical constraint. This unconstrained system does
not define a specific physical theory capable of describing a
cosmological model, as the constraint is the only nontrivial
consequence of Einstein's equation. It is the constrained system
$C_0$ that has physical significance. The operators $\hat{\hat q}$
and $\hat{\hat p}$ (alternatively $\hat q$ and $\hat p$) are
obtained by quantizing the same variables $\varphi$ and
$\pi_\varphi$ but with a different physical meaning, namely as the
classical observables of the physical system $C_0$. Therefore, it
is these operators that are to be identified with the physical
observables of the corresponding quantum system.

If we quantize the Hamiltonian (\ref{K-zero-2}) according to
    \be
    \epsilon\to\hat o_3,~~~~~~~~\varphi=q\to\hat{\hat
    q},~~~~~~~~\pi_\varphi=\pi_q\to\hat{\hat p},
    \label{quantize}
    \end{equation}
we find the quantum Hamiltonian operator
    \be
    \hat h'=\sqrt{\hat{\hat p}^2-\kappa\,
    e^{4\check t\hat o_3}+e^{6\check t\hat o_3}\,V(\hat{\hat q})},
    \label{quantum-Hamiltonian}
    \end{equation}
that yields the dynamics of the evolving Wheeler-DeWitt fields
$\psi_{\check t}$ in terms of their wave functions
$f(q,\epsilon;\check t)$. Using (\ref{wf-20}) and
(\ref{quantum-Hamiltonian}) and rescaling $\check
t\to\epsilon\check t$ for fixed $\epsilon$, we can express the
Schr\"odinger equation (\ref{sch-eq-wf}) in the form
    \be
    i\hbar\frac{\partial}{\partial\check t}\,
    f(q,\epsilon;\epsilon\check t)=\epsilon
    \left[-\hbar^2\frac{\partial^2}{\partial q^2}-\kappa\,
    e^{4\check t}+e^{6\check t}\,V(q)\right]^{1/2}
    f(q,\epsilon;\epsilon\check t).
    \label{sch-eq-wf-2}
    \end{equation}

We can also express the dynamics of $\psi_{\check t}$ by
identifying the form of the Hamiltonian $h'$ appearing in the
Schr\"odinger equation~(\ref{sch-eq-wf-wdw}). In view of the
correspondence between $\hat q$, $\hat p$, $o_3$ and $\hat{\hat
q}$, $\hat{\hat p}$, $\hat o_3$, we have
    \be
    h'=\sqrt{{\hat p}^2-\kappa\,
    e^{4\check t o_3}+e^{6\check t o_3}\,V({\hat q})}.
    \label{quantum-Hamiltonian-wdw}
    \end{equation}
As we discussed in great detail in Section~5, $h'$ does not
generate $\alpha$-translations in the physical Hilbert space
${\cal H}$ of the solutions of the Wheeler-DeWitt equation, and
one cannot identify the $\alpha$ appearing in the Wheeler-DeWitt
equation (\ref{wdw-0-2}) with a time variable (unless one relaxes
the condition that evolution must be unitary with respect to some
inner product on ${\cal H}$.)

The above discussion of the dynamics uses the
Schr\"odinger-picture of quantum mechanics. Having obtained the
basic observables $o_3$, $\hat q$, $\hat p$ and the Hamiltonian
$h'$, we can also study the quantum dynamics in the
Heisenberg-picture. In particular, one has the Heisenberg
equations
    \be
    i\hbar\frac{d}{d\check t}\,\hat q_H(\check t)=
    [\hat q_H(\check t),h'_H(\check t)],~~~~~~~~~~
    i\hbar\frac{d}{d\check t}\,\hat p_H(\check t)=
    [\hat p_H(\check t), h'_H(\check t)],
    \label{hp-wdw}
    \end{equation}
where the subscript $~_H$ denotes the corresponding
Heisenberg-picture observables. Note however that because the
Schr\"odinger-picture Hamiltonian $h'$ is explicitly
time-dependent, it differs from the Heisenberg-picture
Hamiltonian \cite{nova},
    \[h'_H=\sqrt{{\hat p_H}^2-\kappa\,
    e^{4\check t o_3}+e^{6\check t o_3}\,V({\hat q_H})}\neq h'.\]
Here we use the fact that $o_3$ commutes with $h'$, hence the
corresponding Heisenberg-picture operator coincides with $o_3$.

An alternative way of formulating the dynamics is to pursue a
path-integral quantization of the classical system $C_0$. This
amounts to selecting a classical action functional ${\cal S}$ and
postulating the following expression for the propagator of the
theory
    \[\upsilon (q,\epsilon,\check t;q',\epsilon',\check t')=
    \int_{\gamma} {\cal D}_\epsilon q(\check t)~
    e^{{\cal S}[q(\check t),\epsilon(\check t)]}\]
where $\gamma$ stands for the paths in the configuration space
$\R\times\{-1,1\}$ joining the points $(q,\epsilon)$ to
$(q',\epsilon')$, ${\cal D}_\epsilon q$ denotes the `measure' for
the path-integral, and ${\cal S}$ is a classical action functional
for the system $C_0$. Then the dynamics of the Wheeler-DeWitt
fields is given in terms of their wave functions~(\ref{wf})
according to
    \[ f(q,\epsilon;\check t)=
        \sum_{\epsilon'=\pm }\int_{-\infty}^\infty dq'\:
        \upsilon (q,\epsilon,\check t;q',\epsilon',\check t_0)\:
        f(q',\epsilon';\check t_0),\]
where $\check t_0$ is the initial time.

Path-integral quantization of minisuperspace models based on
quantization after imposing the classical constraint has been
studied in the literature, see for example Ref.~\cite{simone}.
Our formulation may be viewed as providing the missing link
between these studies and the more tradition canonical approach
(like ours) that is based on the Wheeler-DeWitt equation.

\section{Discussion and Conclusion}

If we follow Dirac's prescription \cite{dirac} to quantize a
constrained system with a single first-class constraint ${\cal
K}$, then the physical Hilbert space is determined by the quantum
constraint $\hat{\cal K}\psi=0$. For a system with a finite number
of continuous degrees of freedom, this is equivalent to a linear
partial differential equation involving the independent variables
that may be interpreted as the dynamical variables of the
unconstrained (nonphysical) classical system. The Hilbert space
defining the quantum mechanics of the constrained (physical)
system is the solution space ${\cal H}$ of this equation.
Therefore the quantum constraint only determines the vector space
structure of the physical Hilbert space ${\cal H}$. It does not
endow this space with a specific positive-definite inner product.
In this article, we obtained the most general positive-definite
inner product on ${\cal H}$ for a class of constraint equations
called the Klein-Gordon-type equations. We also constructed
explicit unitary maps that related various choices for the inner
product on ${\cal H}$.

Our analysis is based on a simple observation that the states of a
quantum system having ${\cal H}$ as its Hilbert space are not
uniquely determined by the value $\psi(t)$ of a solution $\psi$ of
the corresponding Klein-Gordon-type equation at a given value of
the time-like parameter $t$. The latter requires the knowledge of
a pair of elements of $\tilde{\cal H}$, namely $\psi(t)$ and
$\dot\psi(t)$. This rather trivial observation together with the
demand that the corresponding quantum theory admits a
probabilistic interpretation (which translates into the
mathematical requirement that ${\cal H}$ is to be equipped with a
positive-definite inner product) lead to an explicit construction
of the unique Hilbert space structure on ${\cal H}$.

The developments reported in this paper differ from the
traditional approaches to quantum cosmology
\cite{bryce-1,isham,carlip} as summarized by the following
remarks.
    \begin{enumerate}
    \item It is based solely on the postulates of
quantum mechanics and adheres to Dirac's treatment of quantizing
systems with first class constraints. It relies on the
Wheeler-DeWitt equation, but, in contrast with the conventional
approaches based on the Klein-Gordon inner product or conditional
probabilities, it admits a genuine probabilistic interpretation
and a unitary Schr\"odinger time-evolution.
    \item It does not rely on the properties of the
auxiliary (kinematic) Hilbert space of the unconstrained system.
Neither does it involve a gauge-fixing \cite{woodard} or group
averaging \cite{marolf} scheme. Yet it leads to a class of inner
products on the physical Hilbert space that up to
unitary-equivalence includes all possible positive-definite inner
products, i.e., it yields the unique Hilbert space structure on
${\cal H}$.
    \item It does not involve selecting a time-parameter before
clarifying the kinematic structure of the theory. In a sense, it
decouples the Hilbert space problem from the problem of time. It
further provides invaluable insight as to how one should approach
the problem of selecting a time parameter. It identifies the
latter with the issue of determining a Hamiltonian operator. This
restricts the choice of the possible time-variables. Specifically,
the variable $t$ appearing in the definition of a
nonstationary Klein-Gordon-type equation~(\ref{kgt}), in general,
and the variable $\alpha$ appearing in the Wheeler-DeWitt equation
(\ref{wdw-0-2}), in particular, cannot be identified with a
time-parameter.
    \item It does not identify the solutions of the
Wheeler-DeWitt equation with the wave functions of the universe.
It treats the former as abstract vectors in the physical Hilbert
space and employs the usual definition of a wave function to
identify the wave functions of the universe with the coefficients
of the former in an appropriate position-like basis of the
physical Hilbert space. The quantum theory may be formulated in
terms of these wave functions that are not sensitive to
the choice of the factor-ordering prescription used to specify the
Wheeler-DeWitt equation. This leads to a resolution of the
factor-ordering problem in the kinematic level that seems to apply
generally.
    \item The fact that it involves using a particular choice for a
time-like parameter in the construction of an inner product on the
Hilbert space may be viewed as an indication that it suffers from
a multiple-choice problem \cite{isham} already in the kinematic
level. This is actually not true, because any other choice would
give rise to the same Hilbert space structure on ${\cal H}$. This
is manifestly seen in the formulation of the theory based on the
wave functions. What is essential is the existence of a time-like
variable among the independent variables appearing in the
Wheeler-DeWitt equation which is guaranteed by the Lorentzian
signature of the DeWitt supermetric  \cite{bryce-1} on the
minisuperspace.
    \end{enumerate}

The solutions of the Wheeler-DeWitt equation are the abstract
state vectors in the Hilbert space. Constructing a complete set of
explicit solutions of this equation provides a specific
realization of the Hilbert space. Because the physical content of
the quantum theory is independent of the choice of a realization,
the explicit form of the solutions does not have any physical
significance unless the formulation of the dynamics (the
determination of an appropriate Hamiltonian operator and the
associated time variable) is linked with the particular form of
the Wheeler-DeWitt equation or its solutions.

In this article we explored the basic kinematical structure of the
theory trying to use as small number of physical assumptions as
possible. This leads to a fairly simple description of the general
kinematic structure of the quantum theory of FRW models coupled to
a real scalar field. It is not difficult to see that a similar
treatment applies to more complicated (spatially homogeneous)
minisuperspace models \cite{misner}. Our treatment of the
kinematical aspects of the theory also provides clues for
elucidating its dynamical structure.

We have also outlined a proposal for formulating the dynamics that
involves identifying the time and the physical observables
respectively with a classical time variable and Hermitian operators
obtained by canonically quantizing the observables of the
constrained classical system. This proposal applies to
more general minisuperspace models provided that they admit a
Hamiltonian formulation \cite{ryan}. Its main disadvantage is
that it suffers from the well-known problems of quantization after
imposing the classical constraint(s) \cite{isham}. In particular,
the multiple-choice problem arises and one must face the nonlocal
character of the resulting quantum Hamiltonian operator
(\ref{quantum-Hamiltonian-wdw}) and the fact that for a closed
universe there is a range of the time-variable $\check t$ for
which this operator fails to be Hermitian.

Among the main advantages of this proposal are that it provides a
link between the two widely used approaches to quantum cosmology
namely quantization before and after imposing the constraints and
that it clarifies the conceptual basis of the path-integral
approaches to quantum cosmology that are based on a Schr\"odinger
description of dynamics such as the one used in
Ref.~\cite{simone}. The investigation of the classical limit of
the theory is similar to the one in ordinary nonrelativistic
quantum mechanics. One can in principle employ the WKB
approximation, the mode expansion, or time-dependent perturbation
theory and use these methods to extract the physical implications
of the theory. However, the issue of the initial condition for the
wave function of the universe requires additional input.

The quantum mechanics of Klein-Gordon-type fields developed in
this article has other areas of application. An obvious example is
relativistic quantum mechanics of (non-self-interacting) massive
scalar fields. We can formulate the quantum mechanics of such
fields by identifying the space of the corresponding Klein-Gordon
fields with the Hilbert space of the theory. As we show in
Ref.~\cite{cqg}, the class of inner products (\ref{inv-inn-prod})
for Klein-Gordon field~(\ref{kg}) includes a subclass of Lorentz
invariant inner products (These are the inner products studied in
the Appendix.) In fact, the analysis of Section~3 shows that
requiring relativistic invariance is actually not as significant
as one might imagine: All the positive-definite inner products,
including the energy inner product \cite{energy-inner-product},
the relativistically invariant inner product originally
constructed by Woodard \cite{woodard} using the gauge-fixing method
and rediscovered by Von~Zuben \cite{zuben}), and the inner products
(\ref{inv-inn-prod}}) that include the preceding two as special
cases are all physically equivalent.\footnote{For an earlier
related work see \cite{RH}.}

The main advantage of the application of our method to
Klein-Gordon fields is that it allows for a straightforward
representation of the quantum mechanics of a Klein-Gordon field
based on the Hilbert space ${\cal H}_\star=L^2(\R^3)\oplus
L^2(\R^3)$. In this representation the Hamiltonian $H_\star$ of
(\ref{H-star}) coincides with the Foldy's Hamiltonian
\cite{foldy-1956}. In a sense, our method may be viewed as a way
of explicitly performing a Foldy-Wouthuysen transformation in the
Schr\"odinger-picture of the first-quantized scalar field theory
\cite{foldy-1956}. The Hamiltonian $h$ that is induced from
$H_\star$ by the unitary operator $U$ of (\ref{UU}) is precisely
the generator of the Poincar\'e group (in its spin-zero
representation) corresponding to time-translations. Our approach
is especially suitable for addressing the problem of constructing
a relativistic position operator and the associated localized
\cite{newton-wigner} and coherent states \cite{klauder}. We leave
a more detailed treatment of these issues for a future
publication.

\section*{Acknowledgment}

This work has been supported by the Turkish Academy of Sciences in
the framework of the Young Researcher Award Program
(EA-T$\ddot{\rm U}$BA-GEB$\dot{\rm I}$P/2001-1-1). The useful
discussions with Tekin Dereli and Varga Kalantarov are
acknowledged.

%\newpage
\section*{Appendix}

In Ref.~\cite{cqg}, we identified a class of positive-definite
inner products on the solution space of a Klein-Gordon field in a
Minkowski space that were invariant under Lorentz transformations.
Here we shall consider a generalization of these inner products
for a general stationary Klein-Gordon-type field and present a
derivation of the general form of the corresponding observables.
The results are directly relevant to the issue of constructing and
exploring relativistic position operators and localized states.

Consider the two-parameter family of the operators:
    \be
    L_\pm=\frac{1}{2}\,(c_+^2\pm c_-^2)D^{1/2},
    \label{L-rel}
    \end{equation}
where $c_\pm$ are positive real numbers having the dimension of
$\sqrt t$. Comparing this equation with (\ref{L=}), we see that
(\ref{L-rel}) corresponds to the following choice for the
coefficients $a_{\pm,n}$:
    \be
    a_{\pm,n}=c_\pm e^{i\theta_{\pm,n}}\sqrt{\omega_n},
    \label{eq-star}
    \end{equation}
where $\theta_{\pm,n}\in[0,2\pi)$ are still arbitrary. In view of
(\ref{eq-star}), the operators ${\cal A}_\pm$ of (\ref{eq31}) have
the form
    \be
    {\cal A}_\pm= c_\pm u_\pm D^{1/4},
    \label{eq40}
    \end{equation}
where $u_\pm:\tilde{\cal H}\to\tilde{\cal H}$ are unitary
operators
    \be
    u_\pm=\sum_n e^{i\theta_{\pm,n}}|\phi_n\kt\br\phi_n|
    \label{u-pm}
    \end{equation}
that commute with $D$.\footnote{If some or all the eigenvalues
$\omega_n^2$ of $D$ are degenerate, the operators $u_\pm$ will
have the form $\sum_n\sum_{a,b=1}^{d_n}(u_{\pm,n})_{ab}
|\phi_{n,a}\kt\br\phi_{n,b}|$ where $d_n$ is the multiplicity of
$\omega_n^2$, $a,b$ are degeneracy labels, and $(u_{\pm,n})_{ab}$
are the entries of an arbitrary unitary $d_n\times d_n$ matrix
$u_{\pm,n}$.} In effect, (\ref{L-rel}) is equivalent to requiring
that ${\cal A}_\pm$ have the form (\ref{eq40}) for a pair of
unitary operators $u_\pm$ commuting with $D$.

The observables (\ref{o-psi}) of $q_{L_\pm}$ corresponding to the
choice (\ref{L-rel}) are determined by substituting (\ref{eq40})
in (\ref{J-K=}). This yields
    \be
    J_\pm=\frac{1}{2}\,D^{-1/4}\left[
    O_1\pm \left(\frac{c_+}{c_-}\right)
    {\cal O}\right]D^{1/4},~~~~~~
    K_\pm=\frac{1}{2}\,D^{-1/4}\left[
    \left(\frac{c_-}{c_+}\right)
    {\cal O}^\dagger\pm O_2\right]D^{1/4},
    \label{J-K}
    \end{equation}
where
    \be
    O_1:=u_+^\dagger\tilde O_1 u_+,~~~~~~~
    O_2:=u_-^\dagger\tilde O_2 u_-,~~~~~~~
    {\cal O}:=u_+^\dagger\tilde{\cal O} u_-.
    \label{OOO}
    \end{equation}
As seen from these equations, the effect of the operators $u_\pm$
is a simple redefinition of the observables $O_\star$ of $Q_\star$
which corresponds to changing $\tilde O_1\to O_1,\tilde O_2\to
O_2,$ and $\tilde{\cal O}\to{\cal O}$. This means that for the
cases that $L_\pm$ are given by (\ref{L-rel}), the observables $o$
of $q_{L_\pm}$ have the general form (\ref{o-psi}) where $J_\pm$
and $K_\pm$ are determined (according to  (\ref{J-K})) by three
operators $O_1,O_2,{\cal O}:\tilde{\cal H}\to\tilde{\cal H}$ with
$O_1$ and $O_2$ being Hermitian. Similarly, we can show that the
matrix elements $\cbr\psi,o\phi\ckt$ of $o$ are given by
(\ref{matrix-element}) with
    \be
    {\cal J}_\pm=\frac{c_+^2}{2}\,D^{1/4}\left[
    O_1\pm \left(\frac{c_-}{c_+}\right)
    {\cal O}\right]D^{1/4},~~~~~~
    {\cal K}_\pm=\frac{c_-^2}{2}\,D^{1/4}\left[
    \left(\frac{c_+}{c_-}\right)
    {\cal O}^\dagger\pm O_2\right]D^{1/4}.
    \label{cur-J-K}
    \end{equation}
Inserting these equation in (\ref{matrix-element}) leads to
    \bea
    \cbr\psi,o\phi\ckt&=&\frac{c_+^2}{4}\mbox{\Large$\{$}
    \br\psi_0|D^{1/4}[O_1+r({\cal O}+{\cal O}^\dagger)+r^2O_2]
    D^{1/4}\phi_0\kt+\nn\\
    &&\hspace{9mm}
    \br\dot\psi_0|D^{-1/4}[O_1-r({\cal O}+{\cal O}^\dagger)+r^2O_2]
    D^{-1/4}\dot\phi_0\kt+\nn\\
    &&\hspace{9mm}i\mbox{\Large$($}
    \br\psi_0|D^{1/4}[O_1-r({\cal O}-{\cal O}^\dagger)-r^2O_2]
    D^{-1/4}\dot\phi_0\kt+\nn\\
    &&\hspace{9mm}
    \br\dot\psi_0|D^{-1/4}[-O_1-r({\cal O}-{\cal O}^\dagger)+r^2O_2]
    D^{1/4}\phi_0\kt\mbox{\Large$)$}\mbox{\Large$\}$},
    \label{matrix-element-special}
    \eea
where $r:=c_-/c_+$.

%\newpage

\ed
\begin{thebibliography}{99}
\bibitem{bryce-1} B.\ S.\ DeWitt, Phys.\ Rev.\ {\bf 160}, 1113
(1967).
\bibitem{unruh-wald} W.~G.~Unruh and R.~M.~Wald, Phys.\ Rev.~D
{\bf 40}, 2598 (1989).
\bibitem{isham} C.\ J.\ Isham, in {\em Integrable Systems, Quantum
Groups, and Quantum Field Theories,} edited by L.\ A.\ Ibort and
M.\ A.\ Ropdriguez (Kluwer, Dordrecht, 1993);\\
K.\ Kuch\'ar, in {\em Proceedings of the 4th Canadian Conference
on Relativity and Relativistic Astrophysics}, edited by
G.~Kunstatter, D.~Vincent, and J.~Williams (World Scientific,
Singapore, 1992).
\bibitem{carlip} S.~Carlip, Rep.\ Prog.\ Phys.\ {\bf 64}, 885
(2001); see also \cite{page,wiltshire}.
\bibitem{cqg} A. Mostafazadeh, Class.\ Quantum Grav.\ {\bf 20}, 155
(2003).
\bibitem{misner} C.~W.~Misner, in {\em Magin without Magic: John
Archibald Wheeler,} edited by J.~Klauder (Freeman, San Fransisco,
1972).
\bibitem{kaup-vitello} D.~J.~Kaup and A.~P.~Vitello, Phys.\ Rev.~D
{\bf 9}, 1648 (1974).
\bibitem{blyth-isham} W.~F.~Blyth and C.~J.~Isham, Phys.\ Rev.~D
{\bf 11}, 768 (1975).
\bibitem{page} D.\ N.\ Page, in {\em Gravitation: A Banff Summer
Institute,} edited by R.\ Mann and P.\ Wesson (World Scientific,
Singapore, 1991).
\bibitem{wiltshire} D.\ L.\ Wiltshire, in {\em Cosmology: The Physics
of the Universe,} edited by B.\ Robson, N.\ Visvanathan, and W.\
S.\ Woolcock (World Scientific, Singapore, 1996).
\bibitem{hawking-page} S.~W.~Hawking and D.~N.~Page, Nucl.\
Phys.~B {\bf 264}, 185 (1986).
\bibitem{crnkovic-witten} C.~Crnkovic and E.~Witten, in {\em Three
Hundred Years of Gravitation,} edited by S.~W.~Hawking and
W.~Israel (Cambridge University Press, Cambridge, 1987).
\bibitem{reed-simon} M.\ Reed and B.~Simon, {\em Functional
Analysis,} vol.\ I (Academic Press, San Diego, 1980).
\bibitem{fulling} S.~A.~Fulling, {\em Aspects of Quantum Field
Theory in Curved Space-Time} (Cambridge University Press,
Cambridge, 1989).
\bibitem{kretschmer-szymanowski} R.~Kretschmer and L.~Szymanowski,
`Pseudo-Hermiticity in infinite-dimensional Hilbert Spaces,' LANL
arXiv: quant-ph/0305123.
\bibitem{jpa-98} A.~Mostafazadeh, J.\ Phys.\ A: Math.\ Gen.\
{\bf 31}, 7827 (1998).
\bibitem{foldy-1956} K.~M.~Case, Phys.\ Rev.\ {\bf 95}, 1323
(1954);\\
 L.~L.~Foldy, Phys.\ Rev.\ {\bf 102}, 568 (1956).
\bibitem{FV} H.\ Feshbach and F.\ Villars, Rev.\  Mod.\ Phys.\
{\bf 30}, 24 (1958).
\bibitem{p1} A.\ Mostafazadeh, J.\ Math.\ Phys.\ {\bf 43}, 205 (2002).
\bibitem{p2} A.\ Mostafazadeh, J.\ Math.\ Phys.\ {\bf 43}, 2814 (2002).
\bibitem{p4} A.~Mostafazadeh, Nucl.\ Phys.\ B, {\bf 640}, 419 (2002).
\bibitem{p7} A.~Mostafazadeh, J.\ Math.\ Phys.\ {\bf 44}, 974 (2003).
\bibitem{kato} T.\ Kato, {\em Perturbation Theory for Linear
Operators} (Springer, Berlin, 1995).
\bibitem{p8} A.~Mostafazadeh, `Pseudo-Unitary Operators and
Pseudo-Unitary Quantum Dynamics,' LANL ArXiv: math-ph/0302050.
\bibitem{lewis-riesenfeld} H.~R.~Lewis Jr. and W.~B.~Riesenfeld,
J.~Math.~Phys.\ {\bf 10}, 1458 (1969).
\bibitem{nova} A.~Mostafazadeh, {\em Dynamical Invariants,
Adiabatic Approximation, and the Geometric Phase} (Nova Science
Publishers, New York, 2001).
\bibitem{abe} S.~Abe, Phys.\ Rev.\ D {\bf 47}, 718 (1993).
\bibitem{simone} C.~Simeone, Phys.\ Lett.\ A {\bf 310}, 143 (2003).
\bibitem{energy-inner-product} A.~Ashtekar and A.~Magnon, Proc.\
R.\ Soc.\ Lond.\ A.\ {\bf 346}, 375 (1975);\\
P.~Ghose, D.~Home, and M.~N.~S.~Roy, Phys.\ Lett.~A {\bf 183}, 267
(1993).
\bibitem{messiyah} A.~Messiyah, {\em Quantum Mechanics} (Dover,
New York, 1999).
\bibitem{jmp-98} A. Mostafazadeh, J.\ Math.\ Phys.\ {\bf 39},
4499 (1998).
\bibitem{evans} L.~C.~Evans, {\em Partial Differential Equations}
(American Mathematical Society, Providence, 1998).
\bibitem{vilenkin-89} A.~Vilenkin, Phys.\ Rev.~D {\bf 39}, 1116
(1989).
\bibitem{nakahara} T.~Eguchi, P.~B.~Gilkey, and
A.~J.~Hanson,
Phys.\ Rep.\ {\bf 66}, 213 (1980);\\
M.~Nakahara, {\em Geometry, Topology, and Physics} (Adam Hilger,
Bristol, 1990).
\bibitem{dirac} P.~A.~M.~Dirac, {\em Lectures on
Quantum Mechanics} (Yoshiva University Press, New York, 1964).
\bibitem{woodard} P.\ P.\ Woodard, Class.\ Quantum.\ Grav.\
{\bf 10}, 483 (1993).
\bibitem{marolf} D.\ Marolf, Class.\ Quantum Grav.\ {\bf 12},
1199 (1995); ibid arXiv: gr-qc/0011112;\\
A.~Ashtekar, J.~Lewandowski, D.~Marolf, J.~Mour\~ao, and
T.~Thiemann, J.~Math.\ Phys.\ {\bf 36}, 6456 (1995).
\bibitem{ryan} M.~P.~Ryan Jr., J.\ Math.\ Phys.\ {\bf 15},
812 (1974).
\bibitem{zuben} F.~S.~G.~Von~Zuben, J.\ Math.\ Phys.\ {\bf 41},
6093 (2000).
\bibitem{RH} B.~Rosenstein and L.~P.~Horwitz, J.\ Phys.\ A:
Math.\ Gen.\ {\bf 18}, 2115 (1985).
\bibitem{newton-wigner} T.~D.~Newton and E.~P.~Wigner, Rev.\ Mod.\
Phys.\ {\bf 21}, 400 (1949).
\bibitem{klauder} B.~I.~Lev, A.~A.~Semenov, C.~V.~Usenko, and
J.~R.~Klauder, Phys.\ Rev.\ A {\bf 66}, 022115 (2002).
\end{thebibliography}
